\documentclass[a4paper,10pt]{article}
\usepackage[a4paper, portrait, margin=1in]{geometry}
\usepackage[affil-it]{authblk}
\usepackage[table,usenames,dvipsnames]{xcolor}
\usepackage{color}
\usepackage{hyperref}
\usepackage{wrapfig}
\usepackage{graphicx}
\usepackage{amsfonts}
\usepackage[utf8]{inputenc}
\usepackage[english]{babel}
\usepackage{caption}
\usepackage{subcaption}
\usepackage{amsthm}
\usepackage{amsmath}
\usepackage{amssymb}
\usepackage{physics}
\usepackage[utf8]{inputenc}
\usepackage[english]{babel}
 \usepackage{textcomp}
 \usepackage{multirow}
 \usepackage{mathrsfs,amsmath}
\usepackage[square,numbers]{natbib}
\usepackage{soul}
\usepackage{tikz}
\usetikzlibrary{plotmarks}
\usepackage{etoolbox}
\newcommand{\be}{\begin{equation}}
\newcommand{\ee}{\end{equation}}
\usepackage{caption}
\usepackage{graphicx}
\usepackage{subcaption}
\usepackage{tikz}
\usepackage{etoolbox}

\DeclareRobustCommand\fullblack  {\tikz[baseline=-0.6ex]\draw[black,thick] (0,0)--(0.5,0);}
\DeclareRobustCommand\dotted{\tikz[baseline=-0.6ex]\draw[thick,dotted] (0,0)--(0.54,0);}
\DeclareRobustCommand\dashed{\tikz[baseline=-0.6ex]\draw[thick,dashed] (0,0)--(0.54,0);}

\newrobustcmd*{\mycircle}[1]{\tikz{\filldraw[draw=#1,fill=#1] (0,0) circle [radius=0.05cm];}}
\newrobustcmd*{\mytriangle}[1]{\tikz{\filldraw[draw=#1,fill=#1] (0,0)--(0.2cm,0) -- (0.1cm,0.2cm);}}

\title{Evolutional Deep Neural Network}
\author{Yifan Du, Tamer A.\,Zaki\thanks{corresponding author, email: \url{t.zaki@jhu.edu}}}
\affil{Mechanical Engineering, Johns Hopkins University, Baltimore, MD 21218}

\begin{document}
\maketitle

\begin{abstract}

The notion of an Evolutional Deep Neural Network (EDNN) is introduced for the solution of partial differential equations (PDE).  The parameters of the network are trained to represent the initial state of the system only, and are subsequently updated dynamically, without any further training, to provide an accurate prediction of the evolution of the PDE system.  In this framework, the network parameters are treated as functions with respect to the appropriate coordinate and are numerically updated using the governing equations.  By marching the neural network weights in the parameter space, EDNN can predict state-space trajectories that are indefinitely long, which is difficult for other neural network approaches.
Boundary conditions of the PDEs are treated as hard constraints, are embedded into the neural network, and are therefore exactly satisfied throughout the entire solution trajectory. Several applications including the heat equation, the advection equation, the Burgers equation, the Kuramoto Sivashinsky equation and the Navier-Stokes equations are solved to demonstrate the versatility and accuracy of EDNN. 
The application of EDNN to the incompressible Navier-Stokes equation embeds the divergence-free constraint into the network design so that the projection of the momentum equation to solenoidal space is implicitly achieved. The numerical results verify the accuracy of EDNN solutions relative to analytical and benchmark numerical solutions, both for the transient dynamics and statistics of the system.  

\end{abstract}

\section{Introduction}

The capacity to approximate solutions to partial differential equations (PDEs) using neural network has been an exciting area of research. A key challenge remains the prediction of the dynamics over very long times, that far exceed the training horizon over which the network was optimized to represent the solution.  
In this study, an alternative view is adopted whereby the parameters of an Evolution Deep Neural Networks (EDNN) are viewed as functions in the appropriate coordinate and are updated dynamically, or marched, to predict the evolution of the solution to the PDE for any extent of interest. 

Recent machine learning tools, especially deep neural networks, have demonstrated growing success across computational science domains due to their desirable properties. Firstly, a series of universal approximation theorems \cite{hornik1991approximation, cybenko1989approximation, hornik1989multilayer} demonstrate that neural networks can approximate any Borel measurable function on a compact set with arbitrary accuracy provided sufficient number of hidden neurons. This powerful property allows the neural network to approximate any well defined function given enough samples and computational resources. Furthermore,  \cite{barron1993universal} and more recent studies \cite{yarotsky2018optimal, lu2020deep} provide the estimations of convergence rate of approximation error on neural network with respect to its depth and width, which subsequently allow the neural network to be used in scenarios with high requirements of accuracy. Secondly, the development of differentiable programming and automatic differentiation allow efficient and accurate calculation of gradients of neural network functions with respect to inputs and parameters. These back-propagation algorithms enable the neural network to be efficiently optimized for specified objectives. 

The above properties of neural networks have spurred interest in their application for the solution of PDEs. One general classification of such methods is into two classes: The first focuses on directly learning the PDE operator \cite{li2020fourier, lu2019deeponet}. In the Deep Operator Network (DeepONet), the input function can be the initial and/or boundary conditions and parameters of the equation that are mapped to the output which is the solution of the PDE at the target spatio-temporal coordinates. In this approach, the neural network is trained using data that are often generated from independent simulations, and which must span the space of interest. The training of the neural network is therefore predicated on the existence of a large number of solutions that may be computationally expensive to obtain, but once trained the network evaluation is computationally efficient \cite{cai2020deepm, mao2020deepm}.

The second class of methods adopts the neural network as basis function to represent a single solution. The inputs to the network are generally the spatio-temporal coordinates of the PDE, and the outputs are the solution values at the given input coordinates. The neural network is trained by minimizing the PDE residuals and the mismatch in the initial/boundary conditions.  Such approach dates back to \cite{dissanayake1994neural}, where neural networks were used to solve the Poisson equation and the steady heat conduct equation with nonlinear heat generation.  In later studies \cite{lagaris1998artificial, berg2018unified} the boundary conditions were imposed exactly by multiplying the neural network with certain polynomials. In \cite{weinan2018deep}, the PDEs are enforced by minimizing energy functionals instead of equation residuals, which is different from most existing methods.  
In \cite{raissi2019physics}, a unified neural network methodology called physics-informed neural network (PINN) for forward and inverse (data assimilation) problems of time dependent PDEs is developed. PINNs utilize automatic differentiation to evaluate all the derivatives in the differential equations and the gradients in the optimization algorithm.  Since automatic differentiation consists of analytical derivatives of the activation functions applied repeatedly in a chain rule, gradients in PINNs are evaluated efficiently.
The time dependent PDE is realized by minimizing the residuals at selected points in the whole spatio-temporal domain. The cost function has another penalty term on boundary and initial conditions if the PDE problem is forward, and a penalty term on observations for inverse data assimilation problems.  A schematic representation of the structure and training of PINN is shown in figures \ref{structure_PINN} and \ref{physical_PINN}. The PINN represents the spatio-temporal solution of a PDE as a single neural network, where the behavior in all of space and time is amalgamated in the neural network weights. 
As a result, the causality implicit in the temporal evolution that is inherent to most time dependent PDEs cannot be explicitly specified in PINNs. In addition, the neural network complexity and the dimension of the optimization space grow as the time horizon increases.  As a result, PINNs become computationally expensive for long-time predictions, which motivated the development of time-parallel PINNs \cite{MENG2020} and high-order time-discrete PINNs  (e.g.\,Runge-Kutta 500 \citep{raissi2019physics}).
Nonetheless, for applications to long-time multiscale problems such as chaotic turbulent flows, the storage requirements and complexity of the optimization can become prohibitive.
It is also important to note that the solution of PDEs using PINNs relies on a training, or optimization procedure, where the loss function is a balance between equation residuals and initial/boundary data, and the relative weighting of the two elements as well as the time horizon can frustrate the optimization algorithm \cite{wang2020understanding}. 

In the present effort, a new framework of solving time dependent PDEs, which we term evolutional deep neural network (EDNN), is introduced and demonstrated. The spatial dependence of the solution is represented by the neural network, while the time evolution is realized by evolving, or marching, in the neural network parameter space. Various time dependent PDEs are solved using EDNN as examples to demonstrate its capabilities. In Section \ref{Methodology}, the method of network parameter marching is described in detail, accompanied with a method to embed various constraints into the neural network including boundary conditions and divergence-free constraints for Navier-Stokes equations. In Section \ref{Numerical results} several time dependent PDEs are solved with the newly established EDNN. Various properties of EDNN including temporal and spatial convergence, and long-time predictions are investigated. Conclusions are summarized in section \ref{Conclusions}.

\section{Methodology}\label{Methodology}
\label{Methodology}

Consider a time dependent general nonlinear partial differential equation,  
\be\label{EPDE}
    \frac{\partial \boldsymbol{u}}{\partial t} - \mathcal{N}_{\boldsymbol{x}}(\boldsymbol{u}) = 0, \qquad \boldsymbol{x}\in\Omega\subset\mathbb{R}^d
\ee
where $\boldsymbol{u}(\boldsymbol{x},t) = (u_1, u_2,...,u_m)$ is a vector function on both space and time, the vector $\boldsymbol{x} = (x_1, x_2, ..., x_d)$ contains spatial coordinates, and $\mathcal{N}_{\boldsymbol{x}}$ is a nonlinear differential operator.  
In conventional PINNs, a deep neural network representing the whole time-space solution is trained as shown in figures \ref{structure_PINN} and \ref{physical_PINN}.  For larger time horizons, the network complexity must scale accordingly both in terms of its size and also in terms of training cost which involves optimization of the network parameters. Thus, for very long time horizons, the computational complexity increases appreciably and parallel-in-time algorithms are needed \cite{MENG2020}. The PINN structure is also not designed for making predictions beyond the training horizon, or forecasting. In other words, given a trained PINN for a specific time window, further training is required if the solution is required beyond the original horizon.  

Another approach that aims to evolve the solution of the PDE is reservoir computing \citep{pathak2018model}, where the network inputs and outputs are the solutions at two successive time steps, and the network is thus trained to learn the increment.  In this respect, the governing equations are learned from training data rather than explicitly enforced. 

Here a different approach is introduced: the neural network represents the solution in space only and at a single instant in time, rather than the solution over the entire spatio-temporal domain. Predictions are then made by evolving the initial neural network using the governing equation (\ref{EPDE}). This new framework of using neural network to solve PDEs is called Evolutional deep neural network (EDNN). A schematic of the structure of EDNN and its solution domain are shown in figures \ref{structure_EDNN} and \ref{physical_EDNN}. In this method, the neural network size need only be sufficient to represent the spatial solution at one time step, yet the network has the capacity to generate the solution for indefinitely long times since its parameters are updated dynamically, or marched, using the governing equations in order to forecast the solution. This method is equivalent to discretizing equation (\ref{EPDE}) using neural network on space and numerical marching in time.  It should be noted that the same approach is applicable in any marching dimension, for example along the streamwise coordinate in boundary-layer flows or solving for time-dependent fluid particle positions in Lagrangian formulations  of fluid mechanics.  A key consideration in this new framework is that boundary conditions are no longer enforced through training; instead they must be strictly enforced during the evolution.  

\begin{figure}\label{Schematics_structure} 
    \centering
    \begin{subfigure}{.38\linewidth}\label{structure_PINN}
    \includegraphics[width=0.95\textwidth]{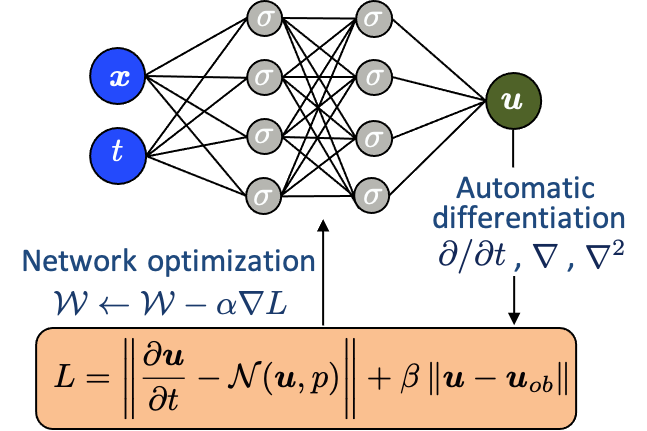}
    \caption{PINN}
    \label{structure_PINN} 
    \end{subfigure}%
    \begin{subfigure}{.60\linewidth}\label{structure_EDNN}
    \centering
    \includegraphics[width=0.95\textwidth]{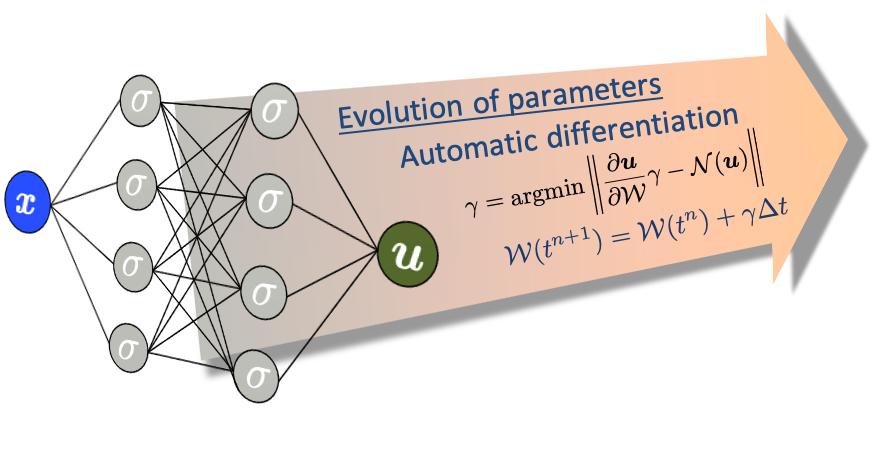}
    \caption{EDNN}
    \label{structure_EDNN}
    \end{subfigure}
    \caption{Schematic representation of PINN and EDNN. (a) PINNs are trained to minimize a cost function comprised of equation residual and data over space and time. (b) The evolution of EDNN, where the network is updated with a direction $\gamma$ calculated from the PDE. The update of neural network parameters represents the evolution of the solution. }
    \label{Schematics_structure}
\end{figure}

\begin{figure}\label{Schematics_physical} 
    \centering
    \begin{subfigure}{.49\linewidth}\label{physical_PINN} 
    \centering
    \includegraphics[width=1.0\textwidth]{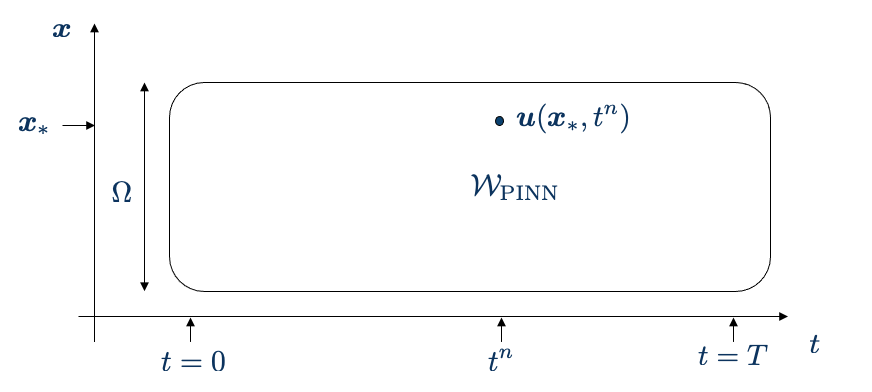}
    \caption{PINN}
    \label{physical_PINN}
    \end{subfigure}%
    \begin{subfigure}{.49\linewidth}\label{physical_EDNN} 
    \centering
    \includegraphics[width=1.0\textwidth]{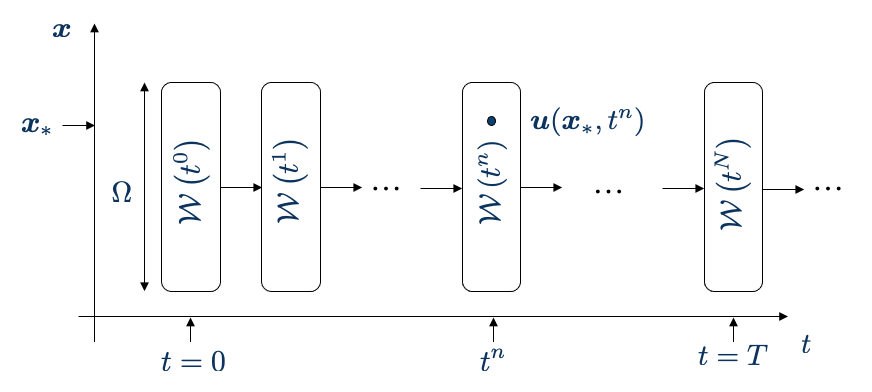}
    \caption{EDNN}
    \label{physical_EDNN} 
    \end{subfigure}
    \caption{Physical domains of PINN and EDNN.  (a) PINN is trained and represents the solution on the whole spatio-temporal domain.  (b) EDNN only represents the solution on space at one time instant. The time evolution of a single network produces the solution trajectory. The network can be evolved indefinitely. }
    \vspace*{-12pt}
    \label{Schematics_physical}
\end{figure}

In section \ref{Method: Evolutional network parameters}, we introduce the detailed algorithm for evolving the neural network parameters.  In section \ref{Method: Embedded constraints}, the approach for enforcing linear constraints on the neural network is discussed, with application to sample boundary conditions. The method of enforcing the divergence-free constraint is also introduced, which will be adopted in the numerical examples using the two-dimensional Navier Stokes equations.

\subsection{Evolutional network parameters}\label{Method: Evolutional network parameters}
Consider a fully connected neural network defined by, 
\be\label{NNit}
\mathbf{g}_{l+1}(\mathbf{g}_l) = \sigma(\mathbf{W}_l\mathbf{g}_l + \mathbf{b}_l),
\ee
where $l\in \{0,1,...,L\}$ is the layer number, $\mathbf{g}_l$ represents the vector containing all neuron elements at the $l^{\textrm{th}}$ layer of the network, $\mathbf{W}_l$ and $\mathbf{b}_l$ represent the kernel and bias between layers $l$ and $l+1$, and $\sigma(\cdot)$ is the activation function acting on a vector element-wise.  Inputs to this neural network are the spatial coordinates of the PDE (\ref{EPDE}), 
$$
\mathbf{g}_0 = \boldsymbol{x} = (x_1, x_2, ..., x_d). 
$$
In this method, we consider the neural network parameters as functions of time $\mathbf{W}_l(t)$ and $\mathbf{b}_l(t)$ so that the whole network is time dependent, and we denote as $\mathcal{W}(t)$ the vector containing all parameters in the neural network.  The output layer $\mathbf{g}_{L+1}$ contains the approximation $\hat{u}$ of the solution to the PDE (\ref{EPDE}), 
$$
\mathbf{g}_{L+1} = \hat{\boldsymbol{u}}\left(\boldsymbol{x}, \mathcal{W}(t)\right) = (\hat{u}_1, \hat{u}_2, ..., \hat{u}_m). 
$$
The dependence of $\hat{\boldsymbol{u}}$ on time is implicitly contained in the neural network parameter $\mathcal{W}(t)$. The time derivative of solution $\hat{u}$ can be calculated according to, $$
\frac{\partial \hat{\boldsymbol{u}}}{\partial t} = \frac{\partial \hat{\boldsymbol{u}}}{\partial \mathcal{W}}\frac{\partial \mathcal{W}}{\partial t}. 
$$
At each time instant, we seek to approximate the time derivative ${\partial \mathcal{W}}/{\partial t}$ by solving, 
\be\label{Wmin}
\frac{\partial \mathcal{W}}{\partial t} = \mathrm{argmin}\mathcal{J}(\gamma), \quad \textrm{where}\, \, \mathcal{J}(\gamma) = \frac{1}{2} \int_{\Omega}\left\Vert\frac{\partial \hat{\boldsymbol{u}}}{\partial \mathcal{W}}\gamma - \mathcal{N}(\hat{\boldsymbol{u}})\right\Vert^2_2\mathrm{d}\boldsymbol{x}, 
\ee
and $\left\Vert \cdot \right\Vert_2$ is the vector 2-norm in $\mathbb{R}^m$. The first-order optimality condition of problem (\ref{Wmin}) yields, 
\be\label{OptCond}
\nabla_{\gamma}\mathcal{J}(\gamma_{opt})  = \left(\int_{\Omega}\frac{\partial \hat{\boldsymbol{u}}}{\partial \mathcal{W}}^{T}\frac{\partial \hat{\boldsymbol{u}}}{\partial \mathcal{W}}\mathrm{d}\boldsymbol{x}\right)\gamma_{opt} -\left(\int_{\Omega}\frac{\partial \hat{\boldsymbol{u}}}{\partial \mathcal{W}}^{T}\mathcal{N}(\hat{\boldsymbol{u}})\mathrm{d}\boldsymbol{x}\right) = 0.
\ee
The optimal solution $\gamma_{opt}$ is approximated by $\hat{\gamma}_{opt}$ which is the solution to, \be\label{OptCondApprox}
\mathbf{J}^{T}\mathbf{J}\hat{\gamma}_{opt} = \mathbf{J}^{T}\mathbf{N}.  
\ee
In the above, $\mathbf{J}$ is the neural network gradient and $\mathbf{N}$ is the PDE operator evaluated at a set of spatial points, 
\be
    \left(\mathbf{J}\right)_{ij} = \frac{\partial \boldsymbol{u}^i}{\partial \mathcal{W}_j}, \quad \, \, \left(\mathbf{N}\right)_{i} = \mathcal{N}(\boldsymbol{u}^i),
\ee
where $i = 1, 2, ..., N_u$ is the index of the collocation point, and $j = 1, 2, ..., N_{\mathcal{W}}$ is the index of the neural network parameter. The elements in $\mathbf{J}$ and $\mathbf{N}$ are calculated through automatic differentiation. It can be shown that as the number of collocation points $N_u \to \infty$, the following holds: 
\be
\frac{1}{N_u}\mathbf{J}^{T}\mathbf{J} \to \frac{1}{\Omega} \int_{\Omega}\frac{\partial \hat{\boldsymbol{u}}}{\partial \mathcal{W}}^{T}\frac{\partial \hat{\boldsymbol{u}}}{\partial \mathcal{W}}\mathrm{d}\boldsymbol{x}, \quad \,\,
\frac{1}{N_u}\mathbf{J}^T\mathbf{N} \to \frac{1}{\Omega} \int_{\Omega}\frac{\partial \hat{\boldsymbol{u}}}{\partial \mathcal{W}}^{T}\mathcal{N}(\hat{\boldsymbol{u}})\mathrm{d}\boldsymbol{x}
\ee

The solution of equation (\ref{OptCondApprox}) is an approximation of the time derivative of $\mathcal{W}$. Two methods that can be utilized to solve (\ref{OptCondApprox}) are direct inversion and optimization.  By using the solution from last time step as initial guess, using optimization method accelerates the calculations compared to direct inversion. Both methods give numerical solutions with satisfactory accuracy. An explicit time discretization scheme can be used to perform time marching, for example forward Euler,  
\be
\frac{ \mathcal{W}^{n+1} - \mathcal{W}^{n}}{\Delta t} =\hat{\gamma}_{opt}^{n}
\ee
where $n$ is the index of time step, and $\Delta t$ is the time step size. For better temporal accuracy, the widely adopted $4^{\textrm{th}}$ order Runge-Kutta scheme can be used,  
\be
    \mathcal{W}^{n+1} = \mathcal{W}^{n} + \left(\frac{1}{6}k_1 + \frac{1}{3}k_2 + \frac{1}{3}k_3 + \frac{1}{6}k_4\right)\Delta t,
\ee
where $k_1$ to $k_4$ are given by, 
\be
    \begin{split}
        k_1 = &\hat{\gamma}_{opt}(\mathcal{W}^{n})\\
        k_2 = &\hat{\gamma}_{opt}(\mathcal{W}^{n}+k_1\frac{\Delta t}{2})\\
        k_3 = &\hat{\gamma}_{opt}(\mathcal{W}^{n}+k_2\frac{\Delta t}{2})\\
        k_4 = &\hat{\gamma}_{opt}(\mathcal{W}^{n}+k_3\Delta t).
    \end{split}
\ee

The initial condition $\mathcal{W}(0) = \mathcal{W}_0$ is evaluated through training the neural network with initial data. The cost, or loss, function of this training is, 
\be\label{cost_initial}
    \mathcal{J}_0(\mathcal{W}^0) = \frac{1}{2}\sum_{i=0}^{N_u} \left\Vert  \hat{\boldsymbol{u}}(\boldsymbol{x}^i, \mathcal{W}^0) - \boldsymbol{u}(\boldsymbol{x}^i,t=t^0)\right\Vert_2^2, 
\ee
where $i = 1, 2, ..., N_u$ represents the index of collocation points. After minimizing (\ref{cost_initial}), the initial condition $\mathcal{W}(0)$ is used in the ordinary differential equation (\ref{Wmin}) to solve for the solution trajectory $\mathcal{W}(t)$. The solution of equation (\ref{EPDE}) then can be calculated at arbitrary time $t$ and space point $\boldsymbol{x}$ by evaluating the neural network using weights $\mathcal{W}(t)$ and input coordinates $\boldsymbol{x}$.

\subsection{Embedded constraints}\label{Method: Embedded constraints}
In this section we discuss a general framework to embed linear constraints into neural networks. Denote by $\mathscr{U}$ and $\mathscr{A}$ Banach spaces, and  $\mathscr{M} \subset \mathscr{U}$ as the neural network function class that is to be constrained. A general linear constraint on $\boldsymbol{u} \in \mathscr{M}$ can be written as follow: 
\be\label{LinearConstraint}
    \mathcal{A}\boldsymbol{u} = 0, \quad \,\,
    \boldsymbol{u} \in \mathscr{M}
\ee
where $\mathcal{A} : \mathscr{U} \to \mathscr{A}$ is a linear operator on $\mathscr{U}$. In most existing deep learning framework for solving PDEs, this constraint is realized by minimizing the following functional, 
\be\label{Jconstraints}
    \mathcal{J}_{A} = \left\Vert \mathcal{A}\boldsymbol{u}\right\Vert_{\mathscr{A}}, \quad \,\,
    \boldsymbol{u} \in \mathscr{M},
\ee
where $\left\Vert \cdot \right\Vert_{\mathscr{A}}$ represents the norm corresponding to space $\mathscr{A}$. Such method only enforces linear constraint (\ref{LinearConstraint}) approximately, and the accuracy of the realization of the constraint depends on the relative weighting between the constraint and other objectives of the training, such as satisfying the governing equations or matching of observation data. 

Instead of minimizing (\ref{Jconstraints}), a general approach is sought to enforce linear constraints exactly. Consider another linear operator $\mathcal{G} : \mathscr{V} \to \mathscr{U}$ as an auxiliary operator for the realization of constraint (\ref{LinearConstraint}). The operator $\mathcal{G}$ satisfies, 
\be\label{AuxG1}
    \mathcal{A}\circ \mathcal{G}(\boldsymbol{v}) = 0, \quad \,\,
    \boldsymbol{v}\in \mathscr{M}'
\ee
where $\boldsymbol{v}$ is the auxiliary neural network function for the realization of constraint $\mathcal{A}$. The function space $\mathscr{M}' \subset \mathscr{V}$ is the neural network function class corresponding to $\boldsymbol{v}$. A sufficient condition of equation (\ref{AuxG1}) is, 
\be\label{AuxG2}
    \mathrm{imag}\left(\mathcal{G}\right) \subseteq \mathrm{ker}\left(\mathcal{A}\right). 
\ee
The problem of enforcing linear constraint (\ref{LinearConstraint}) is thus transformed to the construction of operator $\mathcal{G}$ and the neural network function class $\mathscr{M}'$ that satisfies (\ref{AuxG2}). The newly constructed function \be\label{uGv}
    \hat{\boldsymbol{u}} = \mathcal{G}(\boldsymbol{v}) 
\ee
satisfies the linear constraint $\mathcal{A}(\hat{\boldsymbol{u}}) = 0$. In this way, the linear constraint could be enforced exactly along the solution trajectory. 
Three examples are given below: periodic boundary conditions, homogeneous Dirichlet boundary conditions and a divergence-free condition. 


\subsubsection{Periodic boundary conditions}
The treatment of periodic boundary conditions for the solution of PDE using neural network has been investigated in previous research \cite{yazdani2020systems}. In most of existing methods, input coordinates $\boldsymbol{x}$ are replaced with $\mathrm{sin}(\boldsymbol{x})$ and $\mathrm{cos}(\boldsymbol{x})$ to guarantee periodicity. This method is an example of the general framework discussed here for linear constraints on neural networks. 

Consider a one dimensional interval $\Omega = [0,2\pi]$. The aim is to construct a class of functions that exactly satisfies periodicity on $\Omega$. The linear operator $\mathcal{A}_{p}$ corresponding to periodicity on $\Omega$ is,  
\be
\mathcal{A}_{p}(f) = f(0) - f(2\pi). 
\ee
Choose $\boldsymbol{v} \in \mathscr{M}^{2,1}$ as the auxiliary function, where $\mathscr{M}^{d,q}$ is the neural network function class with input dimension $d$ and output dimension  $q$.  
We construct the auxiliary operator $\mathcal{G}_{p}$ as, 
\be
\mathcal{G}_{p}(\boldsymbol{v})(x) = \boldsymbol{v}\left(\mathrm{sin}\left(x\right), \mathrm{cos}\left(x\right)\right).  
\ee
It can be easily verified that $\mathcal{A}_{p}\circ\mathcal{G}_{p}(\boldsymbol{v}) = 0$.
Examples that involve periodic boundary conditions will be discussed in \S\ref{Res:Hyperbolic}, \S\ref{Res:KS} and \S\ref{Res:INS}.


\subsubsection{Dirichlet boundary conditions}
The homogeneous Dirichlet boundary condition is commonly adopted in the study of PDEs and in applications. A construction of boundary conditions as embedded constraints on a network was achieved in \citep{berg2018unified, luo2020two} by multiplying the network with certain polynomials or by another pre-trained network. Here, a new method for enforcing Dirichlet boundary conditions is introduced. The approach guarantees machine-zero level of error for homogeneous Dirichlet boundary condition on arbitrary geometry and can be trivially extended to higher dimensions.

To state the problem precisely, the constraint operator $\mathcal{A}$ is the trace operator $\mathrm{T}: H^{1}\left(\Omega\right) \to L^2\left(\partial\Omega\right)$, which maps an $H^{1}\left(\Omega\right)$ function to its boundary part. The corresponding auxiliary operator $\mathcal{G}_{\mathrm{T}}$ is not unique. For example, the following construction of $\mathcal{G}_{\mathrm{T}}$ not only guarantees that the homogeneous Dirichlet boundary condition is satisfied, but also provides smoothness properties of the solution, 
\be
\mathcal{G}_{\mathrm{T}}\boldsymbol{v} = \boldsymbol{v} - \int_{\partial \Omega} \frac{\partial \Theta}{\partial \boldsymbol{n}}(\boldsymbol{x},\boldsymbol{y})\boldsymbol{v}(\boldsymbol{y})\mathrm{d}\boldsymbol{y},
\ee
where $\Theta$ is the Green's function of Poisson equation on the domain $\Omega$, and $\boldsymbol{n}$ is the outward unit normal to the boundary.  
The operator $\mathcal{G}_{\mathrm{T}}$ maps any function $f \in H^{1}\left(\Omega\right)$ to a function with zero values on the boundary. However, this construction of $\mathcal{G}_{\mathrm{T}}$ is not ideal. If $\boldsymbol{v}$ is a neural network function, then any single evaluation of
$\boldsymbol{v}(x_0)$ at point $x_0\in\Omega$ requires computing the integral $\int_{\partial \Omega} \frac{\partial \Theta}{\partial n}(\boldsymbol{x}_0,\boldsymbol{y})\boldsymbol{v}(\boldsymbol{y})\mathrm{d}\boldsymbol{y}$, 
which is computationally expensive. Instead, we propose a computationally efficient method to enforce the Dirichlet condition on a domain with arbitrary boundary, which we demonstrate using a two-dimensional example but the construction is easily extended to higher dimensions. 

\begin{figure}\label{DBC}
\centering
\begin{subfigure}{.4\linewidth}
\centering
\includegraphics[width=1.0\textwidth]{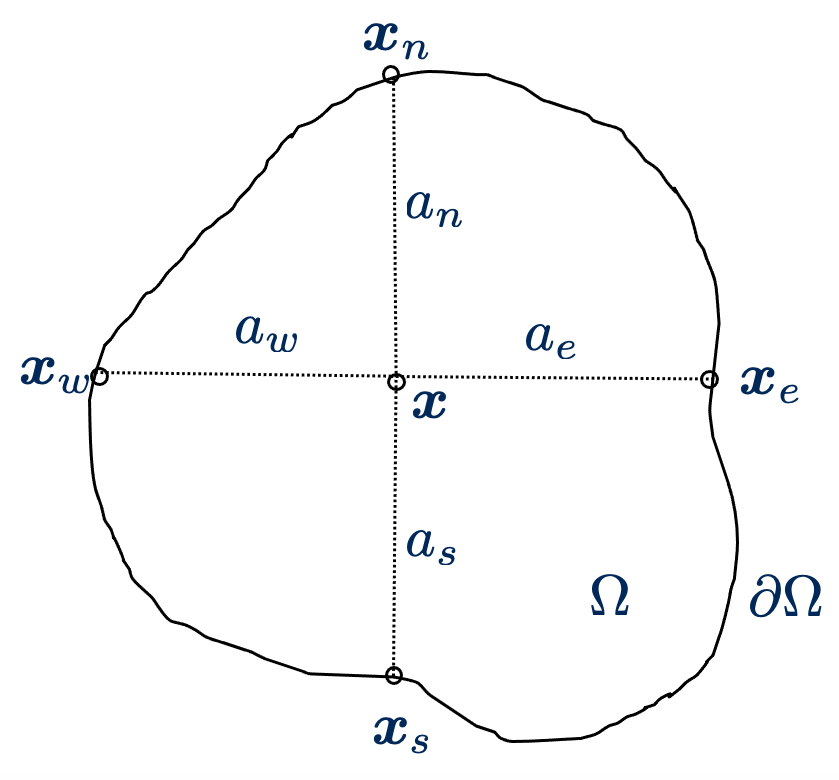}
\caption{Physical domain $\Omega$}\label{DBC_domain}
\end{subfigure}%
\begin{subfigure}{.6\linewidth}
\centering
\includegraphics[width=1.0\textwidth]{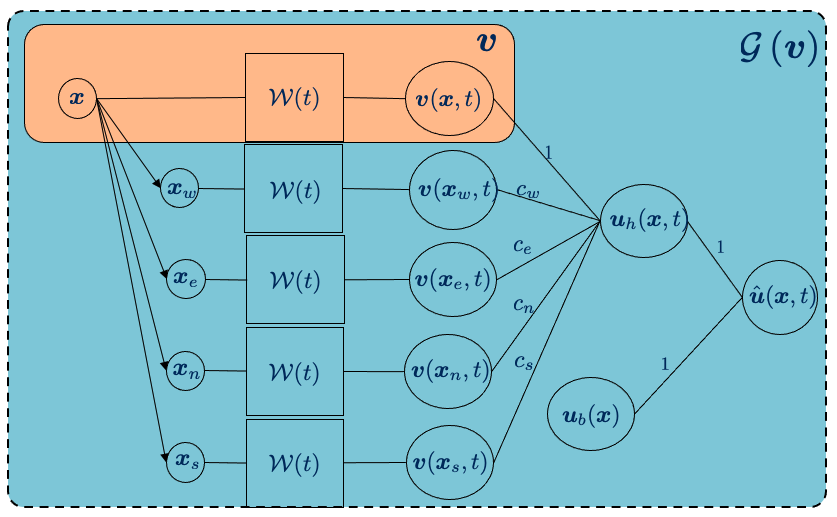}
\caption{Network structure for Dirichlet boundary condition}\label{DBC_NNstructure}
\end{subfigure}
\caption{Schematics for Dirichlet boundary conditions. (a) Geometric quantities including $\boldsymbol{x}_e$, $\boldsymbol{x}_w$, $\boldsymbol{x}_n$, $\boldsymbol{x}_s$ and $a_e$, $a_e$, $a_e$, $a_e$ corresponding to point $\boldsymbol{x}$. (b) Geometrical quantities are used to construct a network that satisfies Dirichlet boundary condition. }
\label{DBC}
\end{figure}

The main idea is that a neural network with homogeneous boundary conditions can be created from an inhomogeneous network by cancelling its boundary values.
For illustration, figure (\ref{DBC_domain}) shows a two-dimensional arbitrary domain $\Omega$.  An arbitrary point in $\Omega$ is denoted $\boldsymbol{x} \in \Omega \subset \mathbb{R}^2$. Horizontal and vertical rays emanating from $\boldsymbol{x}$ intersect the boundary $\partial \Omega$ at $\boldsymbol{x}_e$, $\boldsymbol{x}_w$, $\boldsymbol{x}_n$ and  $\boldsymbol{x}_s$, with corresponding distances $a_e$, $a_w$, $a_n$ and $a_s$, which are all a function of $\boldsymbol{x}$. 
Figure (\ref{DBC_NNstructure}) shows the structure of the neural network that enforces the boundary conditions. The output $\boldsymbol{u}_h(\boldsymbol{x},t)$ is a neural network function with homogeneous Dirichlet boundary conditions, 
\be
\boldsymbol{u}_h(\boldsymbol{x}) = \mathcal{G}_{\mathrm{T}}\boldsymbol{v}(\boldsymbol{x})= \boldsymbol{v}(\boldsymbol{x}) + c_e\boldsymbol{v}(\boldsymbol{x}_e)+ c_w\boldsymbol{v}(\boldsymbol{x}_w)+ c_n\boldsymbol{v}(\boldsymbol{x}_n)+ c_s\boldsymbol{v}(\boldsymbol{x}_s)
\ee
where $\boldsymbol{v}$ is a neural network that has non-zero boundary values. The coefficients $c_e$, $c_w$, $c_n$ and $c_s$ are,  
\be\label{coef_DBC}
    c_e = -\frac{a_w a_n a_s}{a_w a_n a_s + a_e}, \quad 
    c_w = -\frac{a_e a_n a_s}{a_e a_n a_s + a_w}, \quad 
    c_n = -\frac{a_w a_e a_s}{a_w a_e a_s + a_n}, \quad 
    c_s = -\frac{a_w a_e a_n}{a_w a_e a_n + a_s}.
\ee
The choice of the above construction can be motivated by considering, for example, $c_e(a_e, a_w, a_n, a_s)$ which satisfies, 
\be
    c_e(0, a_w, a_n, a_s)=-1, \quad 
    c_e(a_e, 0, a_n, a_s)=c_e(a_e, a_w, 0, a_s)=c_e(a_e, a_w, a_n, 0)=0, \quad 
    \forall a_e, a_w, a_n, a_s. 
\ee
Equation (\ref{coef_DBC}) is one example that satisfies such conditions. 
Once $\boldsymbol{u}_h(\boldsymbol{x},t)$ is obtained, an inhomogeneous Dirichlet condition can be enforced on the network by adding $\boldsymbol{u}_b(\boldsymbol{x})$ which may be an analytical function or provided by another neural network. The final $\hat{\boldsymbol{u}}(\boldsymbol{x},t)$ is the neural network solution that satisfies the Dirichlet boundary conditions. 
Examples where these conditions are applied will be discussed in \S\ref{Res:Parabolic} .

\subsubsection{Divergence free}\label{DivFreeConstraint}

\begin{figure}\label{DivFreeSchematic}
\centering
\includegraphics[width=0.6\textwidth]{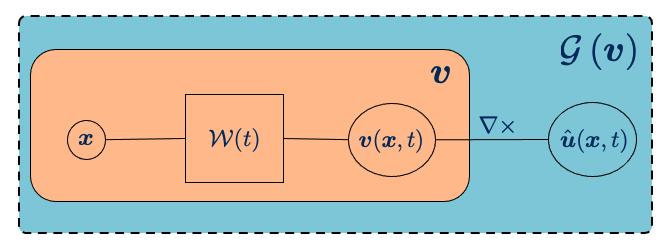}
\caption{Schematics for imposing divergence-free constraint. The shaded regions show he auxiliary network $\boldsymbol{v}$ and $\hat{\boldsymbol{u}}=\mathcal{G}(\boldsymbol{v})$ which satisfies the divergence-free constraint. }
\label{DivFreeSchematic}
\end{figure}

The divergence-free constraint is required for enforcing continuity in incompressible flow fields.  For this constraint, the operator $\mathcal{A}$ is the divergence operator $\mathrm{div}: H^1\left(\Omega;\mathbb{R}^m\right) \to L^2\left(\Omega\right)$. The dimension of the solution domain $dim\left(\Omega\right) = d$ is assumed to be the same as the dimension $m$ of the solution vector. We also denote by  $\mathscr{M}^{d,q}$ the neural network function class with input dimension $d$ and output dimension  $q$. The operator $\mathcal{G}_{\mathrm{div}}$ corresponding to $\mathcal{A}$ can be constructed in different ways depending on $d$: 
\begin{itemize}
  \item $d = 2$: $\boldsymbol{v} \in \mathscr{M}^{2,1} \subset H^2(\Omega,\mathbb{R})$ is the auxiliary neural network function. The auxiliary operator $\mathcal{G}_{div}$ is constructed as: 
  \be
  \mathcal{G}_{div}(\boldsymbol{v})= \left(\begin{matrix}
           ~~{\partial \boldsymbol{v}} / {\partial y} \\
           - {\partial \boldsymbol{v}} / {\partial x}
         \end{matrix}\right), 
  \ee
  In the fluid mechanics context $\boldsymbol{v}$ is the stream function, $\mathcal{G}_{div}$ is the mapping from stream function to velocity field for two-dimensional flow. 
  \item $d = 3$: $\boldsymbol{v} \in \mathscr{M}^{3,3} \subset H^2(\Omega,\mathbb{R}^3)$ is the auxiliary neural network function. The auxiliary operator $\mathcal{G}_{div}$ is constructed as: 
  \be
  \mathcal{G}_{div}(\boldsymbol{v})= \nabla \times \boldsymbol{v}
  \ee
\end{itemize}
A schematic of the above construction is shown in figure \ref{DivFreeSchematic}, and an example of incompressible two-dimensional flow will be presented in \S\ref{Res:INS}.


\section{Numerical results}\label{Numerical results}

In this section, different types of PDEs are evolved using EDNN to demonstrate its capability and accuracy. In \S\ref{Res:Parabolic} the two-dimensional time-dependent heat equation is solved, and the convergence of EDNN to the analytical solution is examined.
In \S\ref{Res:Hyperbolic}, the one-dimensional linear wave equation and inviscid Burgers equation are solved to demonstrate that EDNN is capable to represent transport, including the formation of steep gradients in the nonlinear case.  
In both \S\ref{Res:Parabolic} and \S\ref{Res:Hyperbolic}, we examine the effect of the spatial resolution, and correspondingly the network size, on the accuracy of network prediction.  The influence of the time resolution is discussed in connection with the Kuramoto-Sivashinsky (KS, \S\ref{Res:KS}) and the incompressible Navier-Stokes (NS,\S\ref{Res:INS}) equations, which are nonlinear and contain both advection and diffusion terms.  
The KS test cases (\S\ref{Res:KS}) are used to examine the ability of EDNN to accurately predict the bifurcation of solutions, relative to benchmark spectral discretization.   For the incompressible NS equations (\S\ref{Res:INS}), we compare predictions of the Taylor-Green flow to the analytical solution and provide a comprehensive temporal and spatial resolution test. We also simulate the Kolmogorov flow starting from laminar and turbulent initial conditions.  EDNN can predict the correct trajectory starting from the laminar state, and accurately predict long-time flow statistics in the turbulent regime. In all the following tests we use tanh activation function except for the Burgers equation where we adopt relu activation.  The optimization of the the neural network weights for the representation of initial condition is performed using stochastic gradient descent. 

\subsection{Parabolic equations}\label{Res:Parabolic}
Using the methodology introduced in \S\ref{Methodology}, we solve the two-dimensional heat equation,
\be
    \frac{\partial u }{\partial t} = \nu\left(\frac{\partial^2 u }{\partial x^2}+\frac{\partial^2 u }{\partial y^2}\right), \quad 
    (x,y) \in \Omega = [-\pi, \pi]^2
\ee
with boundary and initial conditions, 
\be
\begin{split}
u(x,y,t=0) = & \,\mathrm{sin}(x)\,\mathrm{sin}(y)\\
 u = & \,0 \quad \,\,\text{on}\,\, \partial \Omega. 
\end{split}
\ee
By appropriate choice of normalization, the heat diffusivity can be set to unity, $\nu=1$. 

The parameters of two tests, denoted 1h and 2h, are provided in Table \ref{heat_params}.  In both cases, the network is comprised of $L=4$ hidden layers, each with $n_L$ neurons.  The smaller number of neurons is adopted for a lower number of collocation points, while the higher value is for a finer spatial resolution.  
Both networks were trained to represent the initial condition until their loss functions reduced by seven orders of magnitude, and subsequently evolved using the algorithm in \S\ref{Methodology}. 

\begin{figure}\label{heat_sol}
    \centering
    \begin{subfigure}{.25\linewidth}
    \centering
    \includegraphics[width=1.0\textwidth]{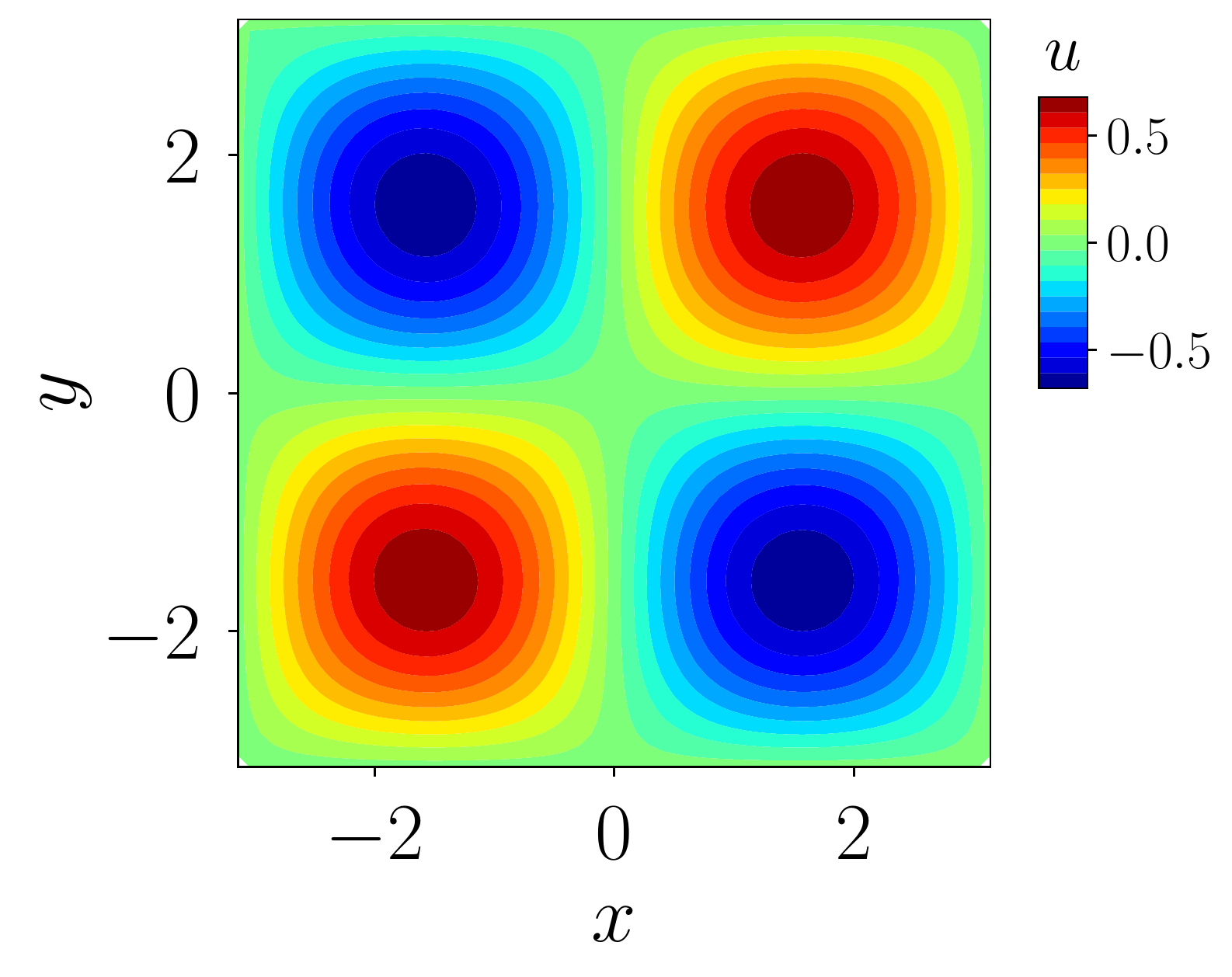}
    \caption{Analytical solution}
    \end{subfigure}%
    \begin{subfigure}{.25\linewidth}
    \centering
    \includegraphics[width=1.0\textwidth]{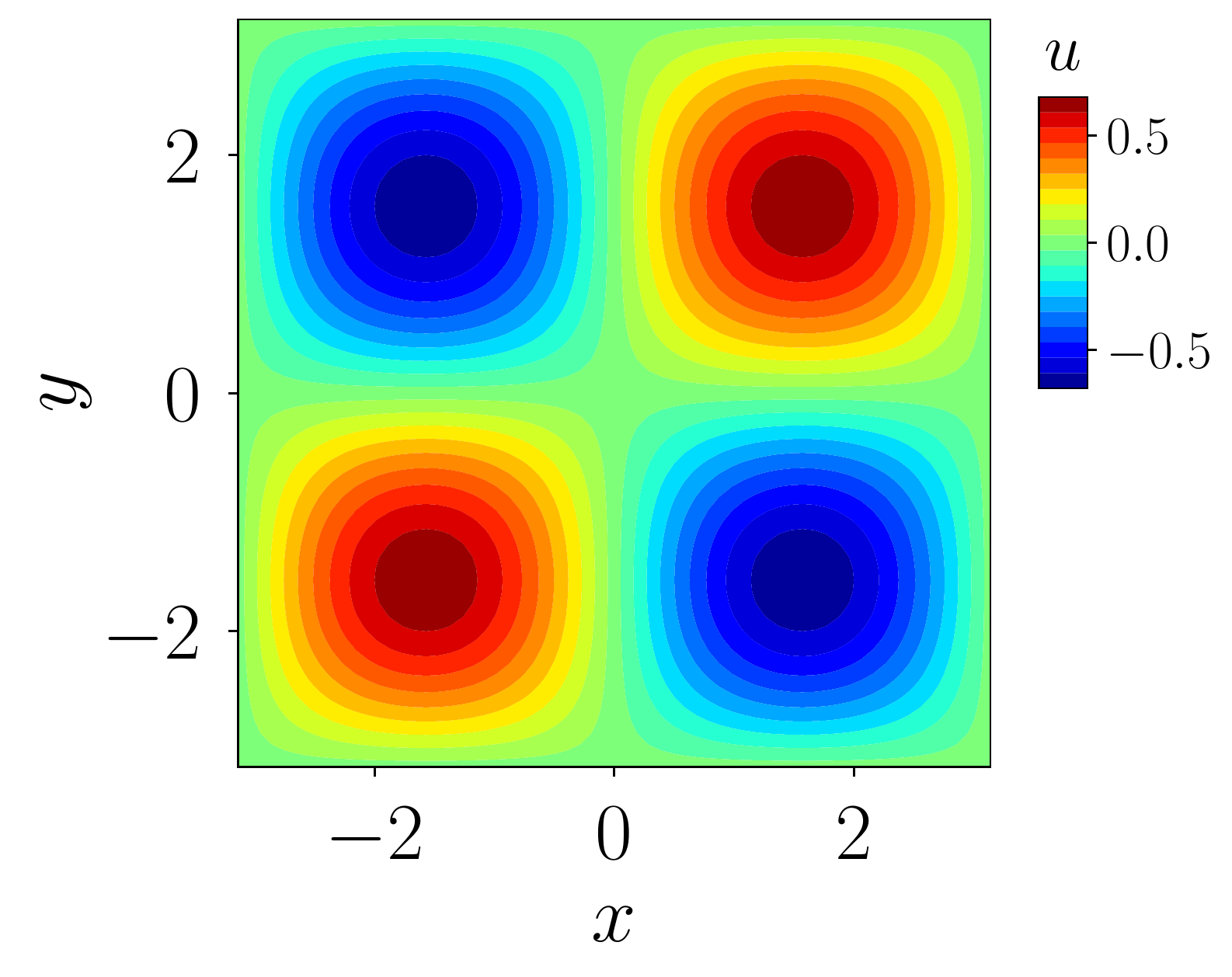}
    \caption{EDNN solution}
    \end{subfigure}%
    \begin{subfigure}{.25\linewidth}
    \centering
    \includegraphics[width=0.85\textwidth]{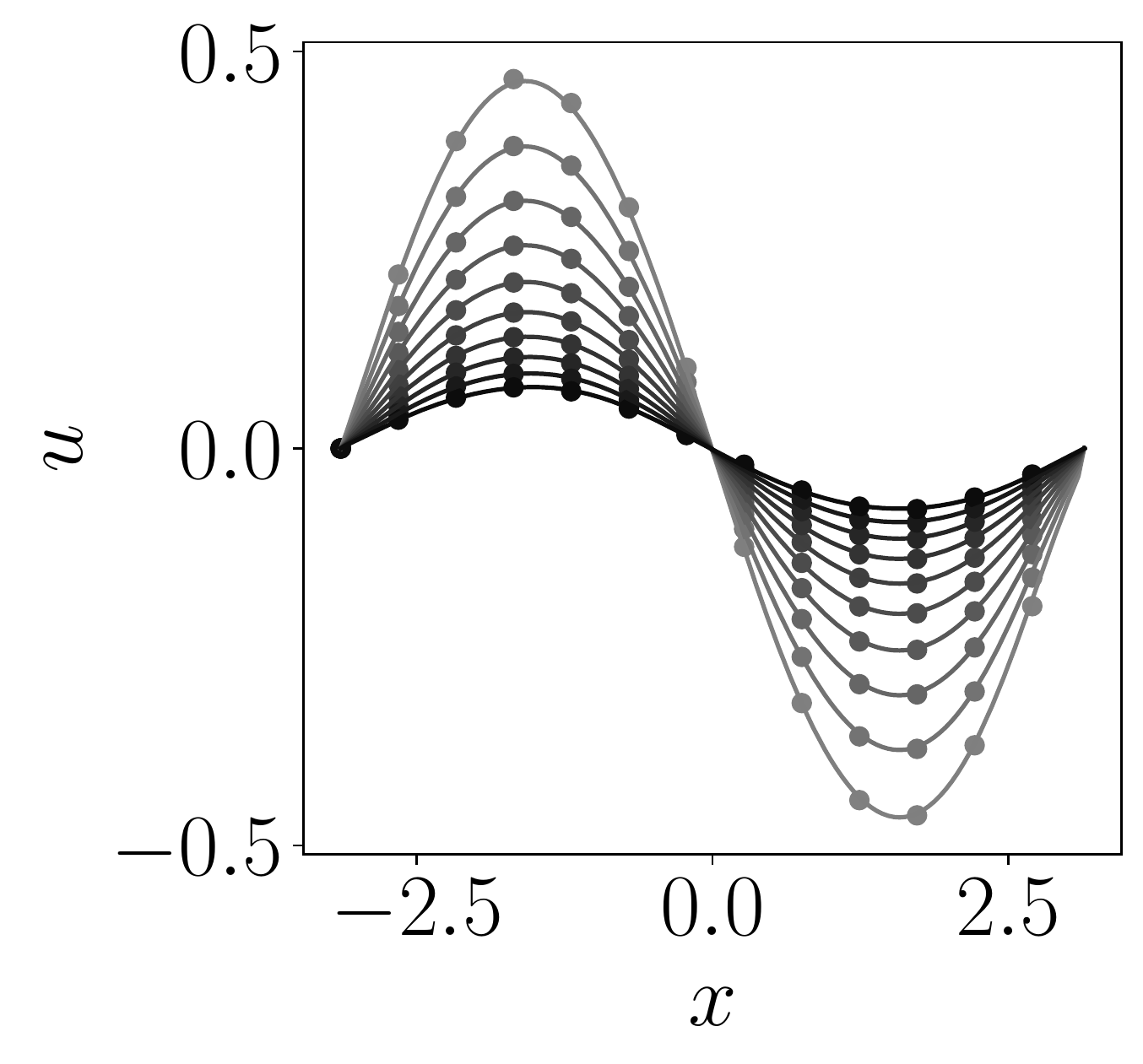}
    \caption{1D comparison}
    \end{subfigure}%
    \begin{subfigure}{.25\linewidth}
    \centering
    \includegraphics[width=0.75\textwidth]{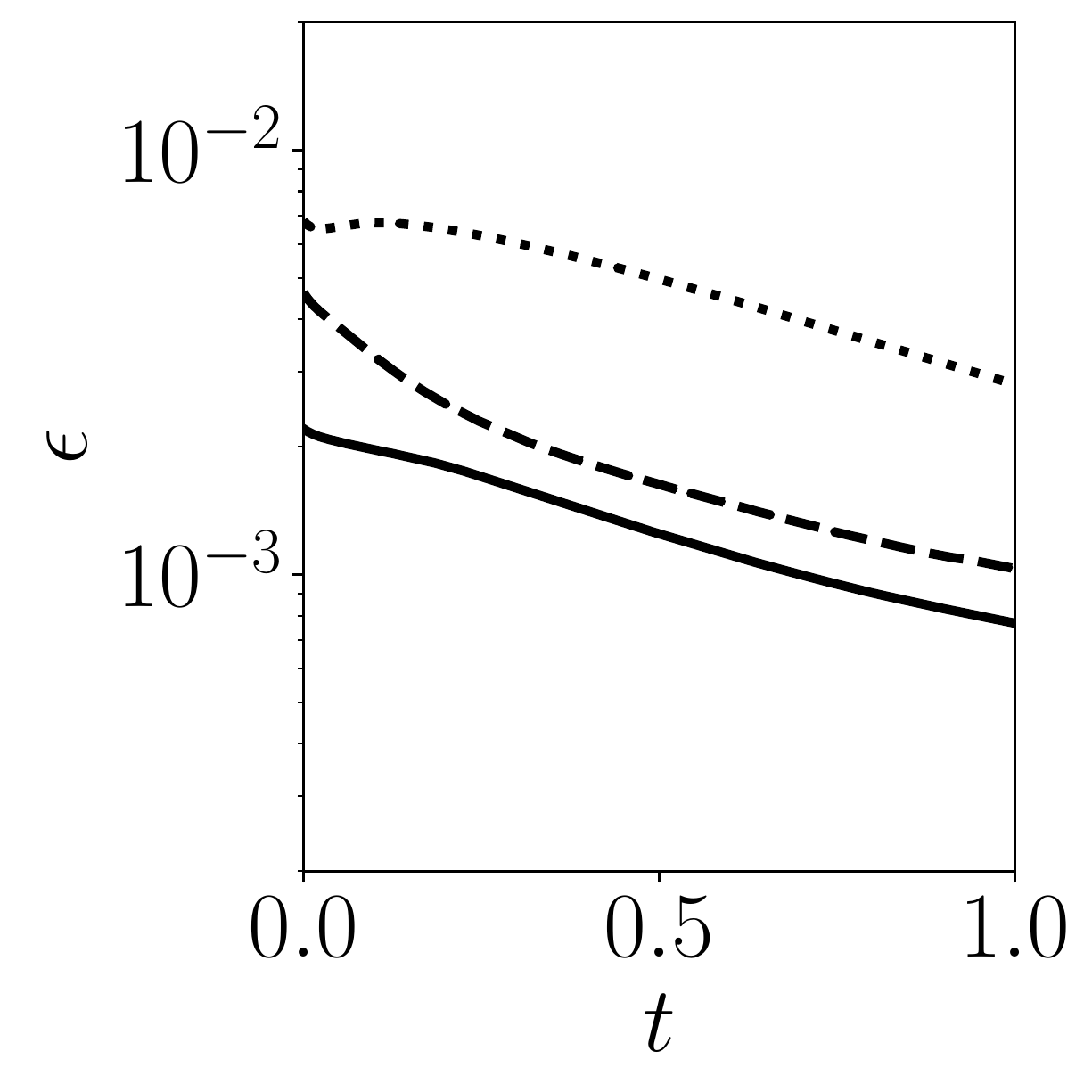}
    \caption{Error versus time}
    \label{heat_sol_error}
    \end{subfigure}%
    \caption{Numerical solution and error evaluation for 2D heat equation using EDNN. (a, b) Contours of True and EDNN solution (case 2h) at $t=0.2$. (c) Comparison between true and EDNN solutions (case 1h) at different times and $y = 1.0$, \mycircle{black!50!gray} \mycircle{black!40!gray} \mycircle{black!40!gray}: true solution,  \fullblack \,EDNN solution. (d) Error of EDNN solution versus time, \dotted \,:case 1h,  \dashed \,: case 2h, \fullblack \,: case 2h with lower initial error. }
\label{heat_sol}
\end{figure}

The predictions of EDNN from case 1h is compared to the analytical solution in figure \ref{heat_sol}. The two-dimensional contours predicted by EDNN display excellent agreement with the the true solution at $t=0.2$.  Panel (c) shows a comparison of the EDNN and true solutions along a horizontal line ($y=1$) at different time instances. Throughout the evolution, the EDNN solution shows good agreement with the analytical result. 
The instantaneous prediction error is evaluated,
\be
    \epsilon = \frac{\left\Vert \hat{u}(t) - u(t) \right\Vert_2}{\left\Vert u(0)\right\Vert_2}
\ee
and is reported in figure \ref{heat_sol_error}. 
The three curves correspond to one simulation using network 1h and two simulations using network 2h.  
In all cases, the errors decay monotonically with respect to time, which indicates that the discretization method we adopt is stable.  
For case 1h, the change in the decay rate at early time can be explained by the initial network not belonging to a typical solution trajectory; it is only trained on the initial data. Once evolved, and after a short transient $(t>0.2)$, the prediction error decays exponentially as expected.

The results from the larger network 2h with spatial refinement of collocation points are more accurate throughout the evolution. For the first of these cases (2h, dashed line), we deliberately started from a finite value of the initial error, associated with training the network to learn the initial condition, that is similar to case 1h.  In this manner, we can highlight the improved accuracy of the predicted solution during its development. Lowering the error associated with the initial state of 2h (solid line) further reduces the error throughout the time history.

%

\begin{table}
\caption{Parameters for linear heat equation calculations using EDNN. }
\begin{center}\label{heat_params}
\begin{tabular}{ ccccccccc } 
 \hline
 Case & $L$ & $n_L$ &  $N_x$ & $N_y$ &  $\Delta t$ &  $\nu \Delta t /\Delta x ^2$\\ 
 \hline
1h & \multirow{2}{*}{$4$}  &  20  &   65 &  65     & $1 \times 10^{-3}$ & $0.10$  \\ 
 2h &                      & $30$ & $129$ & $129$ &        $1 \times 10^{-3}$ & $0.42$ \\
\hline
\end{tabular}
\label{heat_params}
\end{center}
\end{table}

\subsection{Hyperbolic equations}\label{Res:Hyperbolic}
In this section, EDNN is applied to solution of the one-dimensional linear advection equation and the one-dimensional Burgers equation in order to examine its basic properties for a hyperbolic PDE. The linear case is governed by, 
\be\label{1dBurgers}
    \frac{\partial u}{\partial t} + c\frac{\partial u }{\partial x} = 0, \quad\, x \in [-1,1], \,\, c = 1.
\ee
The initial condition is a sine wave,
\be
    u(x,0) = -\mathrm{sin}(\pi x), 
\ee
and periodicity is enforced in the streamwise direction.  
EDNN predictions will be compared to the analytical solution,  
\be
    u=-\mathrm{sin}\left(\pi \left(x-ct \right)\right). 
\ee
The parameters of the calculations are provided in Table \ref{LW_params} (cases 1lw and 2lw).  In both cases, the EDNN architecture is comprised of four layers ($L=4$) each with either 10 (case 1lw) or 20 (case 2lw) neurons.   The number of solution points is increased with the network size, while the timestep is held constant.  

The EDNN prediction (case 2lw) and the analytical solution are plotted superposed in figure \ref{1dLinearWave_sol}, and show good agreement.  The root-mean-squared errors in space $\epsilon$ are plotted as a function of time in panel (b), and demonstrates that the solution trajectories predicted by EDNN maintain very low level of errors. Note that the errors maintain their initial values, inherited from the netowrk representation of the initial condition, and are therefore smaller for the larger network that provides a more accurate representation of the initial field.  In addition, the errors do not amplify in time, but rather oscillate with smaller amplitude as the network size is increased.  This trend should be contrasted to conventional discretizations where, for example, diffusive errors can lead to decay of the solution and an amplification of errors in time. 
\begin{table}
    \caption{Parameters for linear wave equation calculations using EDNN. }
    \begin{center}\label{LW_params}
    \begin{tabular}{ cccccc } 
     \hline
     Case & $L$ & $n_L$ & $N_x$ &   $\Delta t$ \\ 
     \hline
     1lw & \multirow{2}{*}{$4$} & $10$ & $500$ & $1\times 10^{-3}$\\ 
     2lw &                      & $20$ & $1000$ & $1\times 10^{-3}$\\ 
     \hline
     1b  &          4           & $20$ & $1000$ & $1\times 10^{-3}$\\ 
    \hline
    \end{tabular}
    \label{LW_params}
    \end{center}
\end{table}

\begin{figure}\label{1dLinearWave_sol} 
    \centering
    \begin{subfigure}{.25\linewidth}
    \centering
    \includegraphics[width=1.0\textwidth]{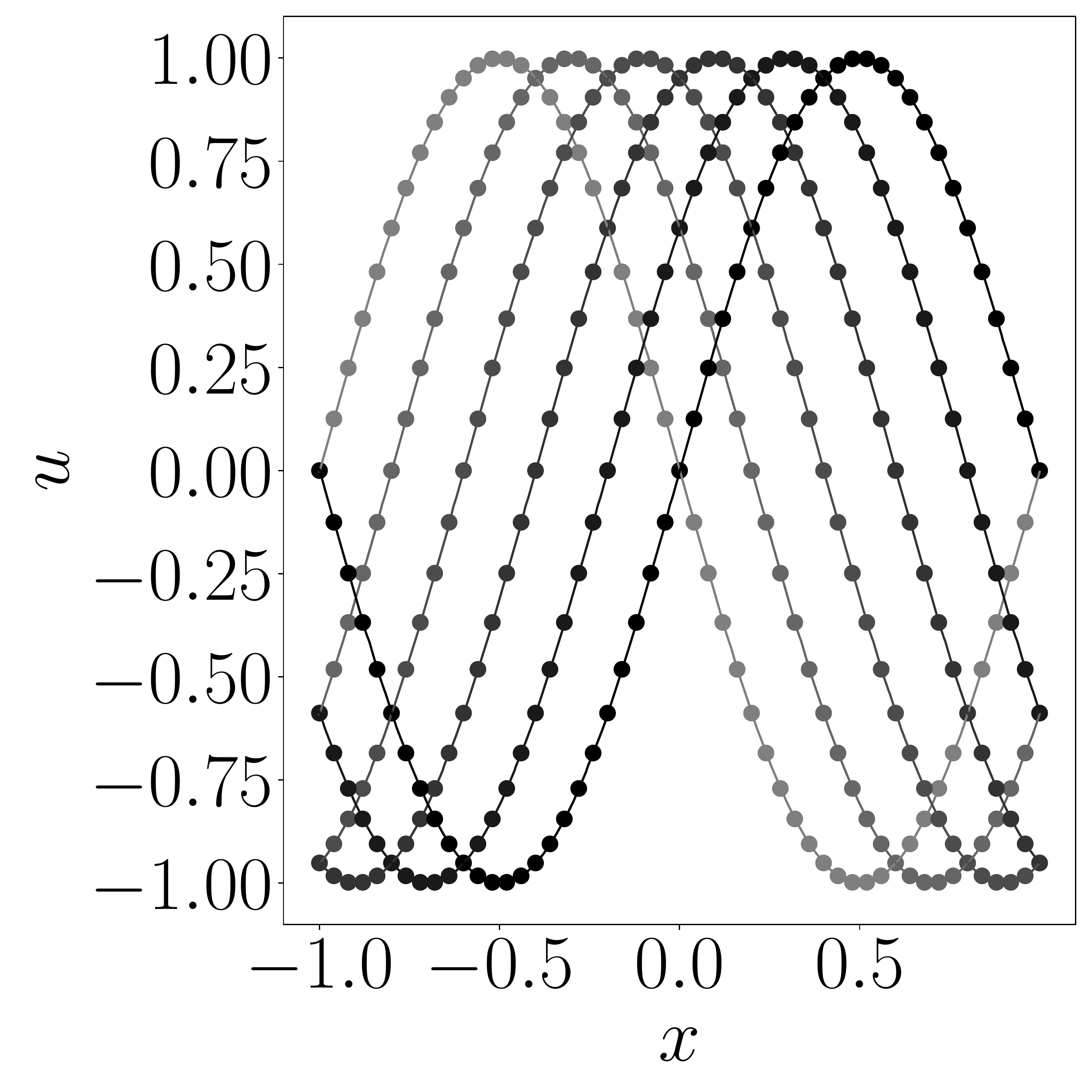}
    \caption{}
    \end{subfigure}%
    \begin{subfigure}{.25\linewidth}
    \centering
    \includegraphics[width=1.0\textwidth]{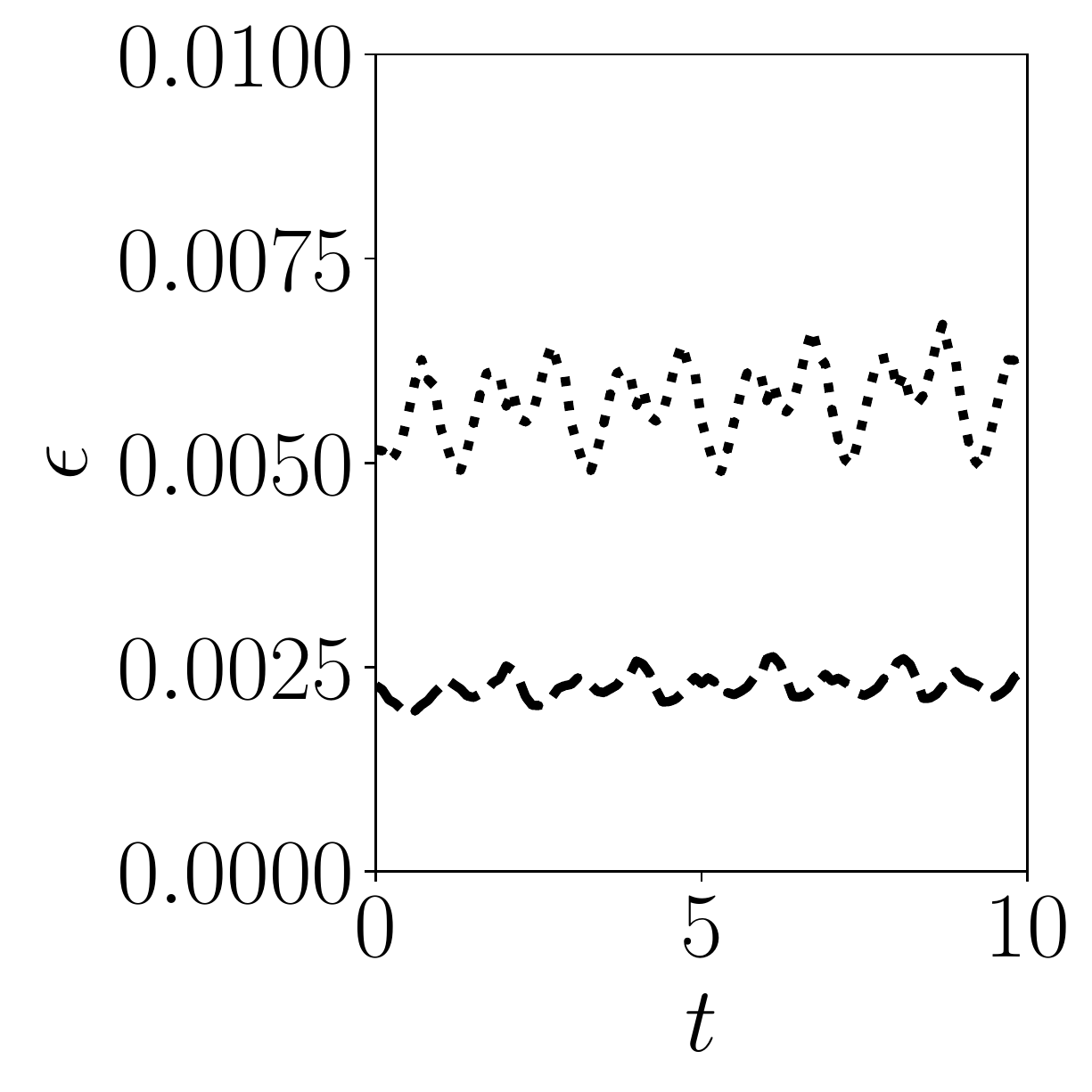}
    \caption{}
    \end{subfigure}
    \caption{Numerical solution of linear wave equation using EDNN. (a) Spatial solution from case 2lw every $0.2$ time units. Symbolds: \mycircle{black!50!gray} \mycircle{black!40!gray} \mycircle{black!40!gray}: true solution,  \fullblack \,EDNN solution. (b) Relative error: \dotted: case 1lw, \dashed: case 2lw}
    \label{1dLinearWave_sol}
\end{figure}

The same EDNN for the linear advection equation can easily be adapted for the non-linear Burgers equation.  The formation of shocks and the capacity of NN to capture them (e.g.\,using different activation functions) is a topic that warrants a separate dedicated effort \cite{mao2020physics}.  For the present scope, one option is to introduce a viscous term to avoid the formation of discontinuities in the solution \cite[see e.g.][]{li2020fourier}; Since we have already simulated the heat equation, here we retain the inviscid form of the Burgers equation and simulate its evolution short of the formation of the N-wave.  We therefore solve, 
\be\label{1dBurgers}
    \frac{\partial u}{\partial t} + u\frac{\partial u }{\partial x} = 0, \quad\, x \in [-1,1]
\ee
with the initial condition,  
\be
    u(x,0) = -\mathrm{sin}(\pi x), 
\ee
with periodic boundary conditions on the given interval $[-1,1]$. 
The analytical solution is given implicitly by the characteristic equation, 
\be\label{1dBurgers_true}
    u=-\mathrm{sin}\left(\pi \left(x-ut \right)\right). 
\ee
This expression is solved using a Newton method to obtain a reference solution.

The parameters of the EDNN used for the Burgers equation is shown in Table \ref{LW_params} (case 1b). The EDNN prediction is compared to the reference solution in figure \ref{1dBurgers_sol} at different stages.  At early times (panel a), the gradient of solution is not appreciable and is therefore resolved and accurately predicted by the network. At the late stages in the development of the N-wave (panel b), the solution develop steep gradient at $x=0$ and becomes nearly discontinuous. The prediction from EDNN continues to accurately capture the reference solution. 


\begin{figure}\label{1dBurgers_sol} 
    \centering
    \begin{subfigure}{.25\linewidth}
    \centering
    \includegraphics[width=1.0\textwidth]{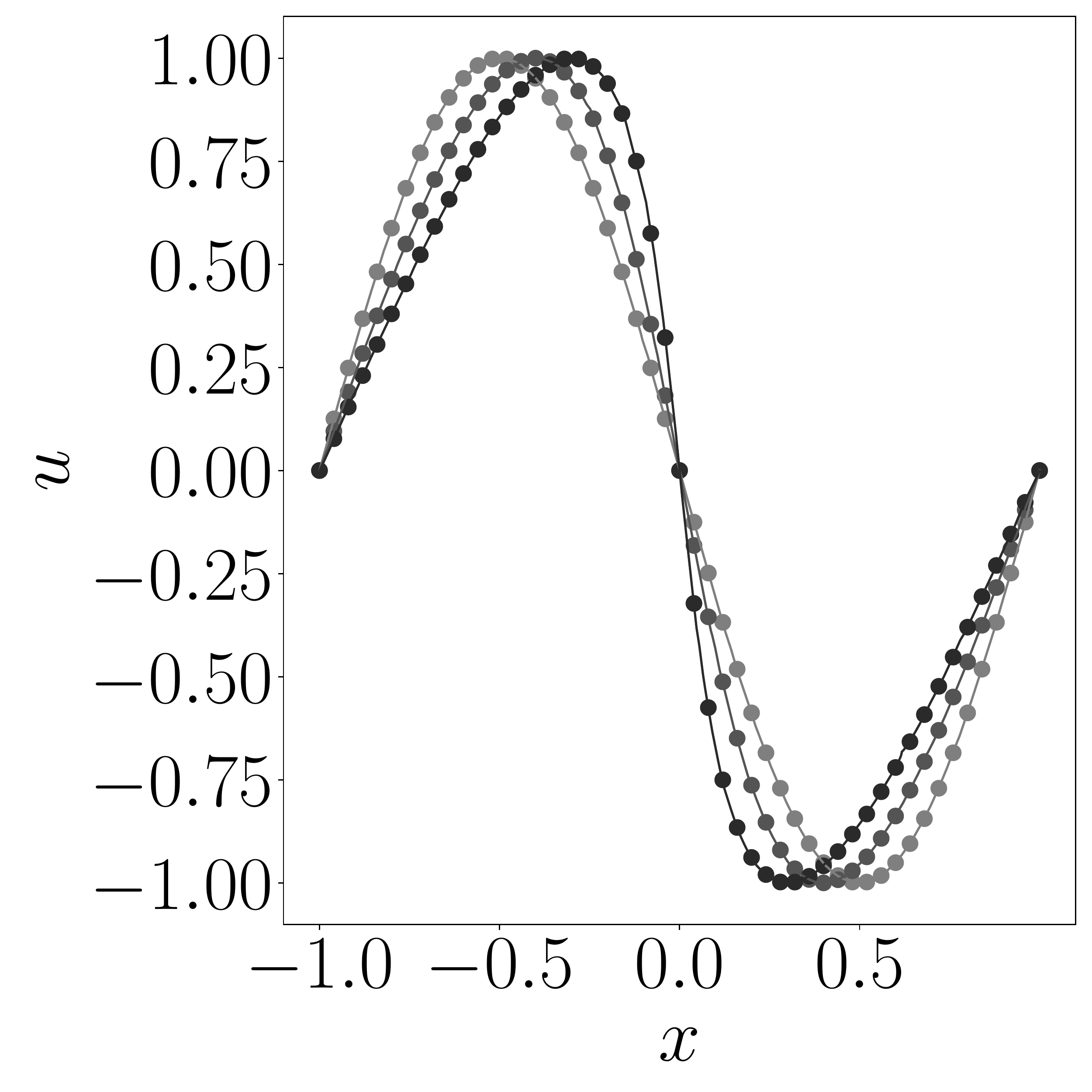}
    \caption{Solution in space at different times}
    \end{subfigure}%
    \begin{subfigure}{.25\linewidth}
    \centering
    \includegraphics[width=1.0\textwidth]{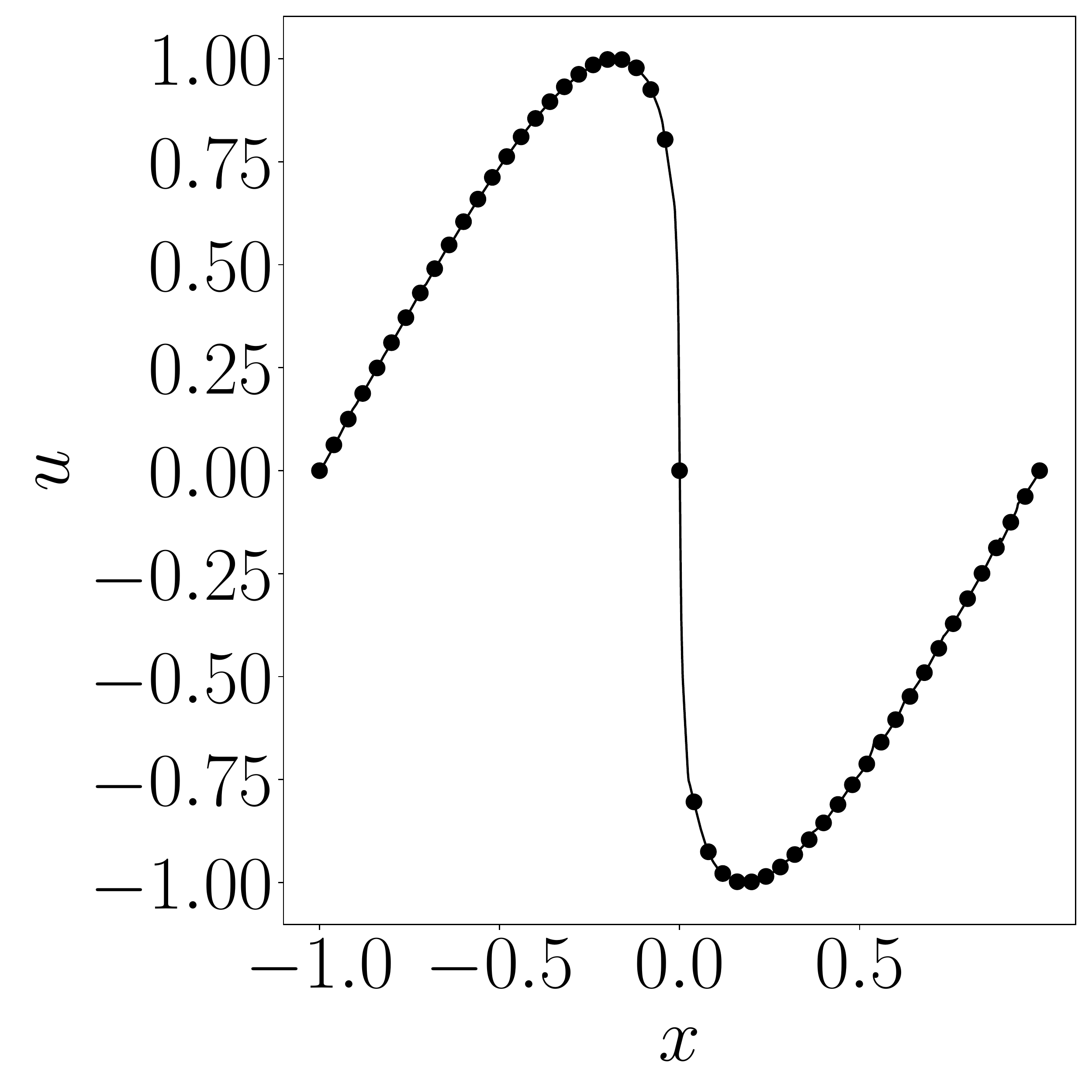}
    \caption{N-wave formation.}
    \end{subfigure}
    \caption{Numerical solution of N-wave formation using EDNN. (a) Solution at $t = \{0.0, 0.1, 0.2\}$. (b) Solution at $t = 0.32$. Symbols: \mycircle{black!50!gray} \mycircle{black!40!gray} \mycircle{black!40!gray}: true solution,  \fullblack \,EDNN solution.}
    \label{1dBurgers_sol}
\end{figure}

\subsection{Kuramoto-Sivashinsky equation}\label{Res:KS}
In this section, the Kuramoto-Sivashinsky (KS) equation is solved using EDNN. The nonlinear $4^{\textrm{th}}$ order PDE, is well known for its bifurcations and chaotic dynamics, and has been subject of extensive numerical study \cite{hyman1986kuramoto, pathak2018model, page2020revealing}.  We will focus on the ability of EDNN to predict bifurcations of the solution, and reserve the discussion of chaotic solutions to simulations of the Kolmogorov flow and its long-time statistics (\S\ref{Res: Kolmogorov flow}).  
We consider the following form of the KS equations, 
\be\label{KS}
    \frac{\partial u}{\partial t} + u\frac{\partial u}{\partial x}+\frac{\partial^2 u}{\partial x^2}+\frac{\partial^4 u}{\partial x^4}=0
\ee
with periodic boundary conditions at the two end points of the domain, and the initial condition,
\be
    u(x,t=0) = - \mathrm{sin}\left(\frac{\pi x}{10}\right),\quad x\in[-10,10]
\ee

The parameters for solving equation (\ref{KS}) using EDNN are provided in Table (\ref{KS_params}). All three cases adopt the same EDNN architecture, with four layers ($L=4$) each with twenty neurons $n_L=20$. The spatial domain is represented by $N_x=1000$ uniformly distributed points, although the method does not impose any restriction on the sampling of the points over the spatial domain which could have been, for example, randomly uniformly distributed. Cases 1k and 2k adopt the same time-step $\Delta t$, and are intended to contrast the accuracy of forward Euler (FE) and Runge-Kutta (RK) time marching schemes for updating the network parameters.  Case 3k also uses RK but with a finer time-step.

\begin{table}
\caption{Parameters for the numerical solution of Kuramoto-Sivashinsky equation using EDNN}
\begin{center}\label{KS_params}
\begin{tabular}{ ccccccc } 
 \hline
Case & $L$ & $n_L$ & $N_x$ &   $\Delta t$ & time discretization\\ 
 \hline
 1k &     &      &        & $1\times 10^{-2}$ & FE\\ 
 2k & $4$ & $20$ & $1000$ & $1\times 10^{-2}$ & RK\\ 
 3k &     &      &        & $1\times 10^{-3}$ & RK\\ 
\hline
\end{tabular}
\label{KS_params}
\end{center}
\end{table}

\begin{figure}\label{KS_sol}
    \centering
    \begin{subfigure}{.33\linewidth}
    \centering
    \includegraphics[width=1.0\textwidth]{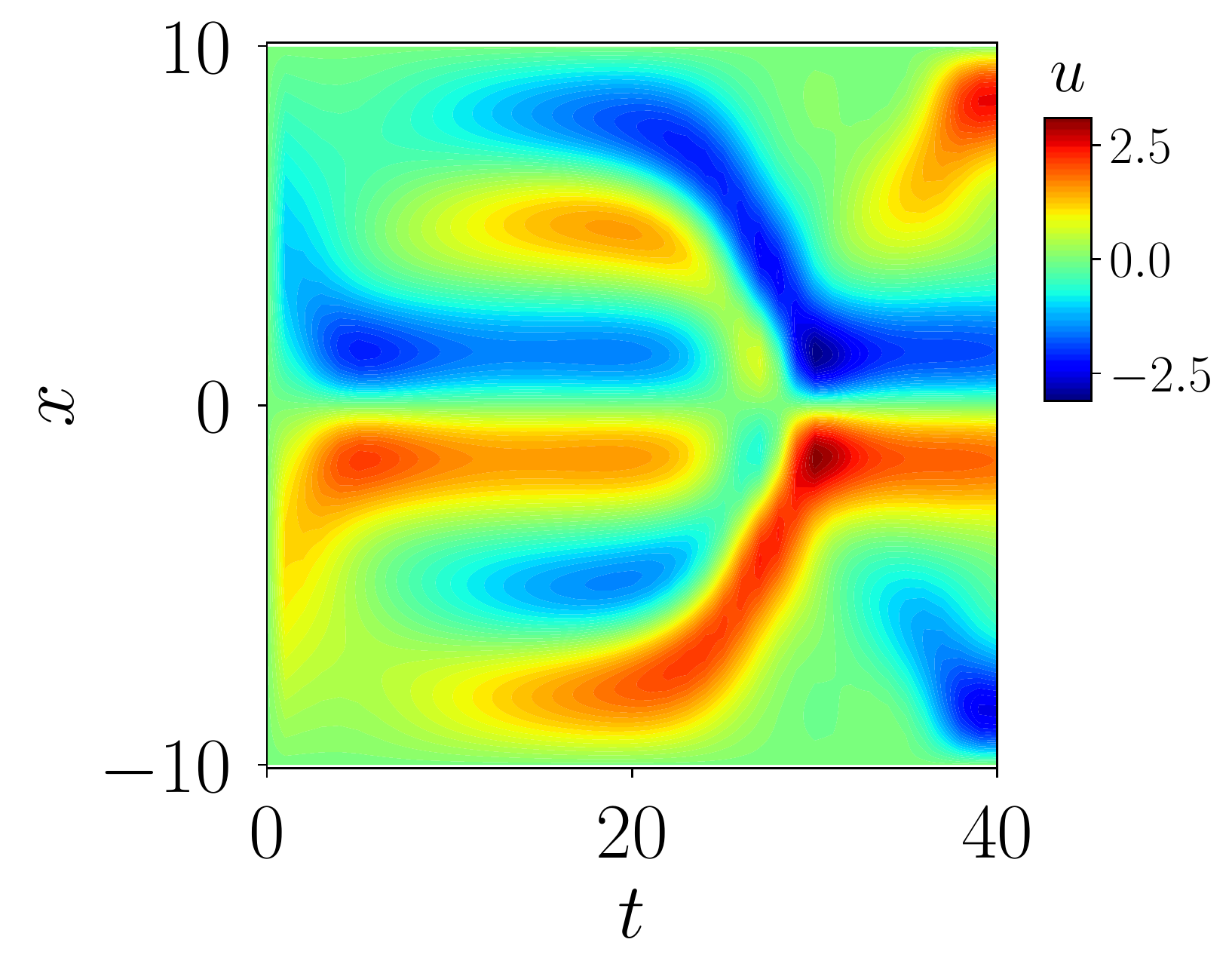}
    \caption{}
    \end{subfigure}%
    \begin{subfigure}{.33\linewidth}
    \centering
    \includegraphics[width=1.0\textwidth]{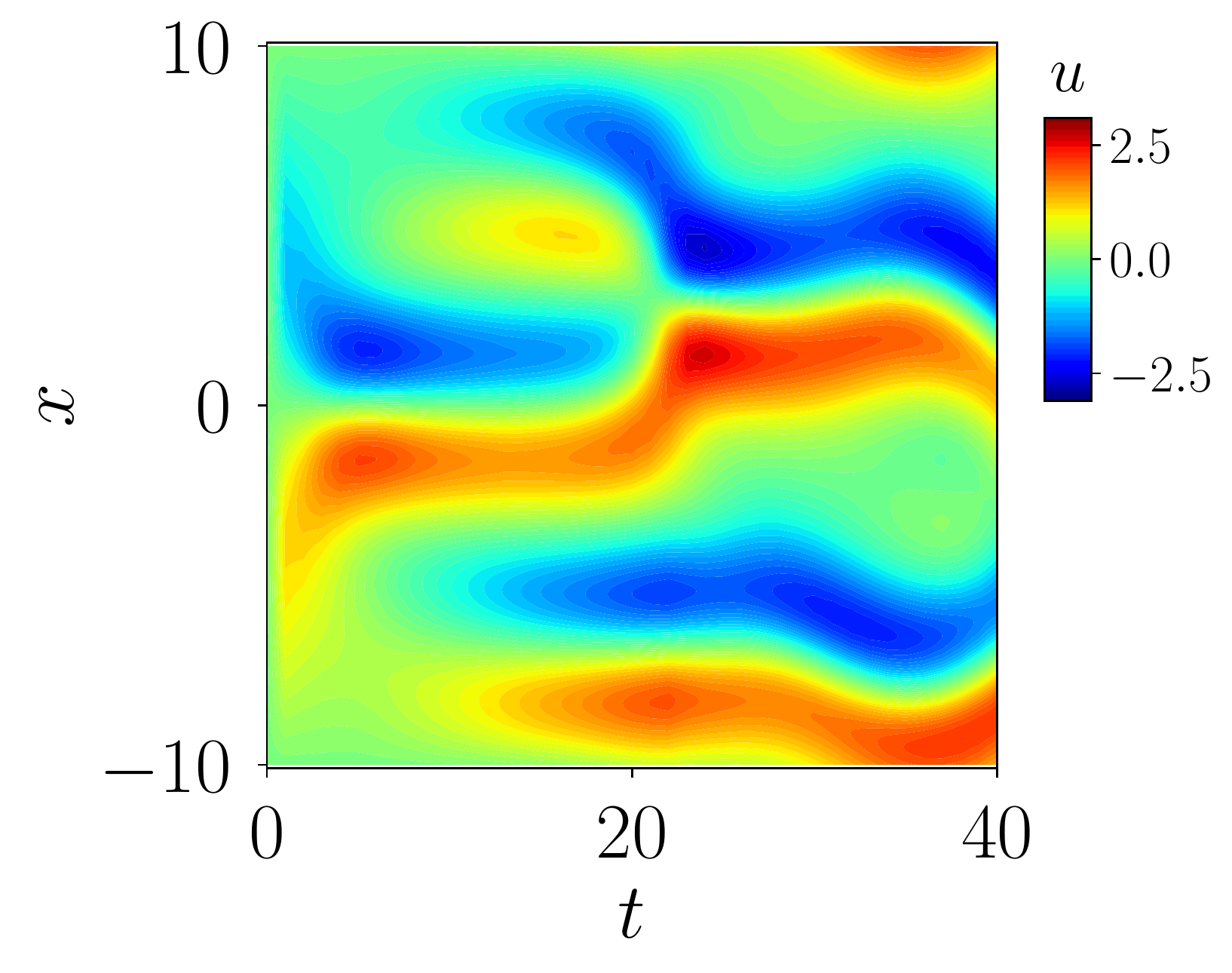}
    \caption{}
    \end{subfigure}%
    \begin{subfigure}{.33\linewidth}
    \centering
    \includegraphics[width=1.0\textwidth]{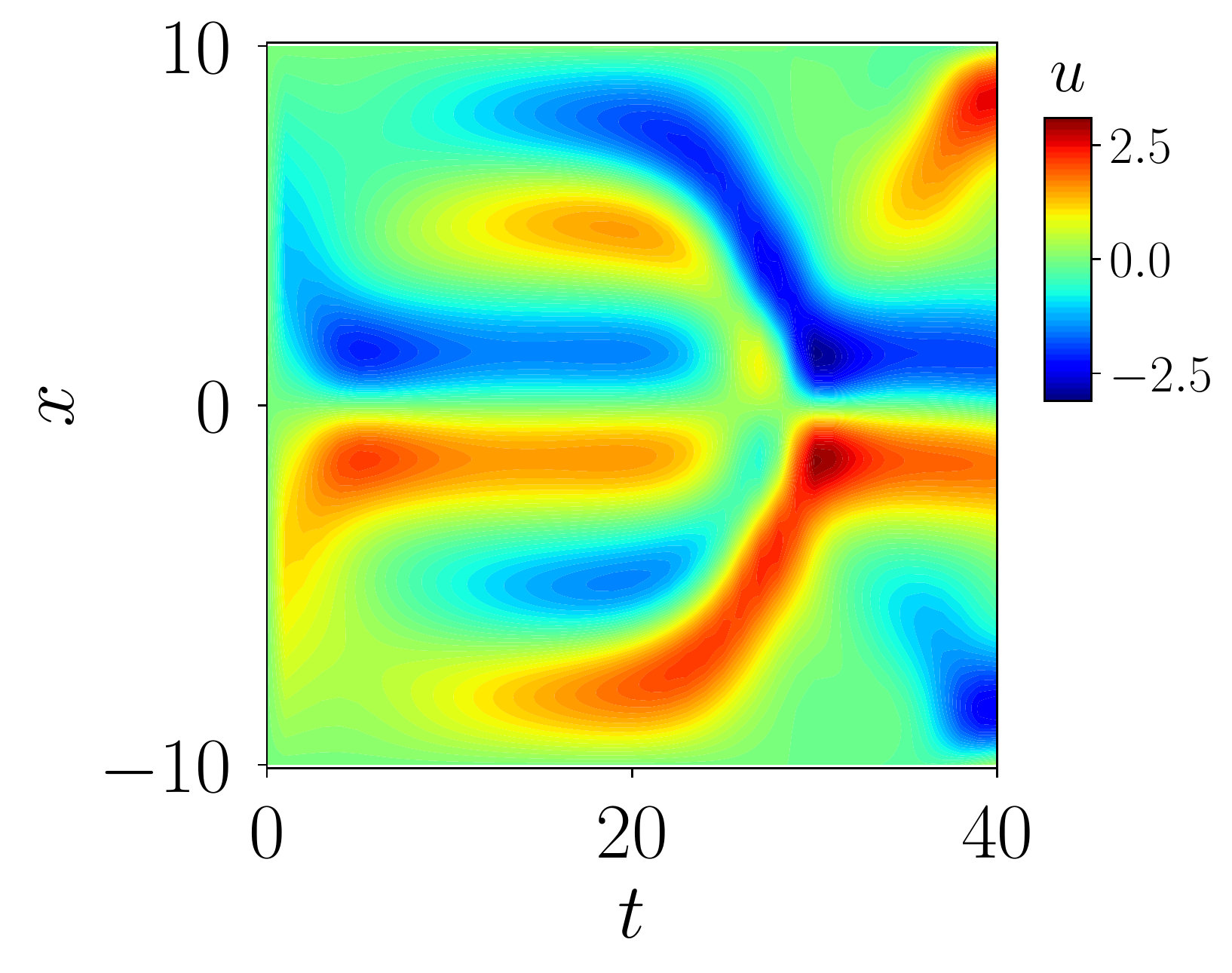}
    \caption{}
    \end{subfigure}
    \caption{Numerical solution of one-dimensional Kuramoto Sivashinsky equation using EDNN. (a): numerical solution from spectral discretization; (b) case 2k; (c) case 3k.}
    \label{KS_sol}
\end{figure}

\begin{figure}\label{KS_err}
    \centering
    \begin{subfigure}{.3\linewidth}
    \centering
    \includegraphics[width=1.0\textwidth]{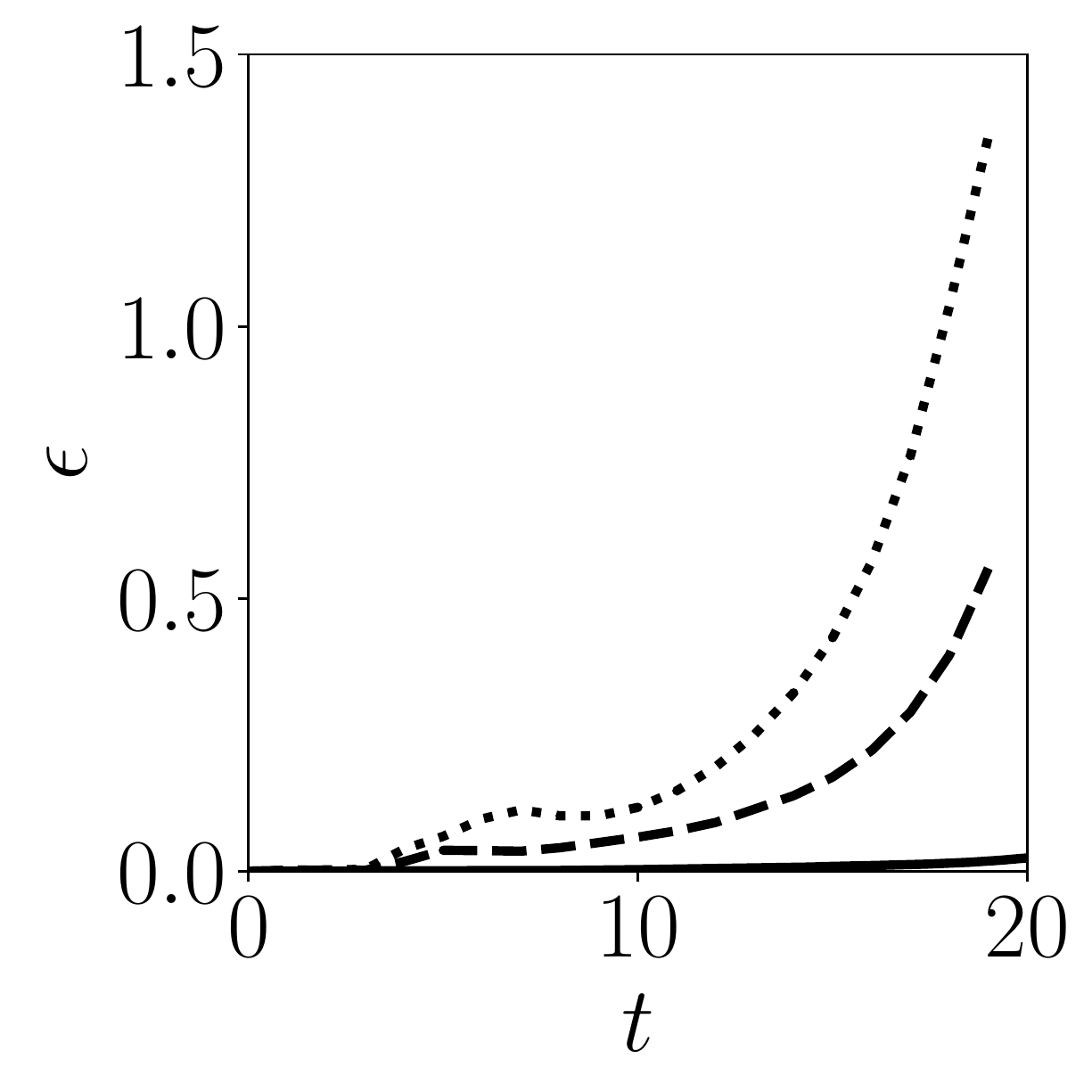}
    \caption{}
    \end{subfigure}%
    \centering
    \begin{subfigure}{.3\linewidth}
    \centering
    \includegraphics[width=1.0\textwidth]{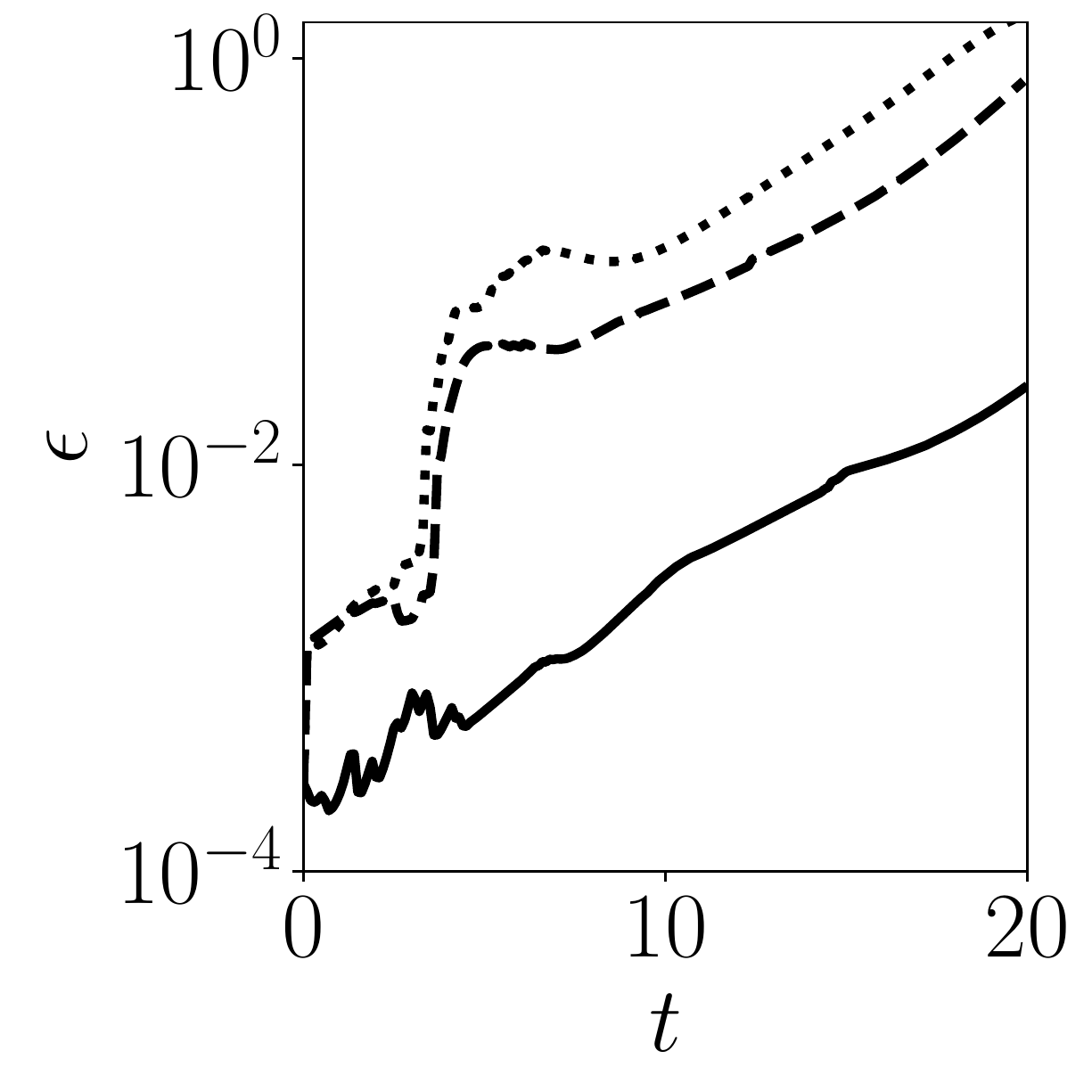}
    \caption{}
    \end{subfigure}
    \caption{Temporal evolution of errors in KS solution using EDNN relative to Fourier spectral method. \dotted \,: case 1k; \dashed \,: case 2k; \fullblack \,: case 3k. Errors $\epsilon$ are reported in (a) linear and (b) logarithmic scale.}
    \label{KS_err}
\end{figure}


Figure (\ref{KS_sol}$a$) shows the behavior of a reference solution, evaluated using a spectral Fourier discretization in space and exponential time differencing $4^\textrm{th}$ order Runge-Kutta method \cite{kassam2005fourth} with $\Delta t = 10^{-3}$ .  Panels (b) and (c) show the predictions from cases 2k and 3k using EDNN. 
The solution of case 2k diverges from the reference spectral solution for two reasons. 
Firstly, the time step size $\Delta t$ in case 2k is large compared to the spectral solution, which introduces large discretization errors in the time stepping. In case 3k, the step size $\Delta t$ is reduced to $10^{-3}$ and the prediction by EDNN shows good agreement with the reference spectral solution. Secondly, the trajectory predicted by solving the KS equation is very sensitive to its initial condition. That initial state is prescribed by training to set the initial state of EDNN, and therefore the initial condition is enforced with finite precision, in this case $O(10^{-3})$ relative error. The initial error is then propagated and magnified through the trajectory of the solution, as in any chaotic dynamical system.

The errors between the reference spectral solution and the three cases listed in table \ref{KS_params} are evaluated,
\be
    \epsilon = \frac{\left\Vert \hat{u}(t) - u(t)\right\Vert_2}{\left\Vert u(0)\right\Vert_2}
\ee
and shown in figure \ref{KS_err}, both in linear and logarithmic scales.  The Euler time advancement of the Network parameters shows the earliest amplification of errors, or divergence of the trajectories predicted by EDNN and the reference spectral solution.  At the same time-step size, the RK time marching has lower error and reducing its time-step size even further delays the amplification of $\epsilon$.  Despite this trend, since the equations are chaotic, even infinitesimally close trajectories will ultimately diverge in forward time at an exponential Lyapunov rate.  Therefore, when plotted in logarithmic scale, the errors all ultimately have the same slope, but the curves are shifted to lower levels for RK time marching and smaller time step.

\subsection{Incompressible Navier-Stokes equations}\label{Res:INS}
In this section we simulate the evolution of the two-dimensional Taylor-Green vortices and of Kolmogorov flow using EDNN. Both cases are governed by the incompressible Navier-Stokes equations, 
\be\label{NS1}
\begin{split}
    \nabla \cdot \boldsymbol{u} &= 0, \\
    \frac{\partial\boldsymbol{u}}{\partial t} + \boldsymbol{u}\cdot\nabla\boldsymbol{u} &= -\nabla P + \nu \nabla^2 \boldsymbol{u} + \boldsymbol{f}, \\
\end{split}
\ee
where $\boldsymbol{u}$ and $P$ represent the velocity and pressure fields, and $\boldsymbol{f}$ represents a body force. 
An alternative form of the equations \cite{temam2001navier, temam1991remark},
\be\label{NS2}
    \frac{\partial\boldsymbol{u}}{\partial t}  = \mathcal{P}\left[-\boldsymbol{u}\cdot\nabla\boldsymbol{u} + \nu \nabla^2 \boldsymbol{u} + \boldsymbol{f}\right]
\ee
replaces the explicit dependence on pressure by introducing $\mathcal{P}$ which is an abstract projection operator from $H^1\left(\Omega\right)$ to its subspace $H^1(\Omega)_{div}$. This form (\ref{NS2}) of the Navier-Stokes equation can be solved directly using EDNN, where the projection operator $\mathcal{P}$ is automatically realized by maintaining a divergence-free solution throughout the time evolution. 

The minimization problem (\ref{Wmin}) corresponding to the Navier-Stokes equations (\ref{NS2}) is, 
\be\label{WminNS1}
    \mathcal{J}_{P}(\gamma) = \frac{1}{2}\int_{\Omega}\left\Vert\frac{\partial \hat{\boldsymbol{u}}}{\partial \mathcal{W}}\gamma - \mathcal{P}\left[-\hat{\boldsymbol{u}}\cdot\nabla\hat{\boldsymbol{u}} + \nu \nabla^2 \hat{\boldsymbol{u}} + \boldsymbol{f}\right]\right\Vert^2_2\mathrm{d}\boldsymbol{x}.
\ee
When the methodology from \S(\ref{DivFreeConstraint}) is adopted to constrain $\hat{\boldsymbol{u}}$ to the solenoidal space, the above cost function can be re-written without the project operator, 
\be\label{WminNS2}
    \mathcal{J}(\gamma) = \frac{1}{2}\int_{\Omega}\left\Vert\frac{\partial \hat{\boldsymbol{u}}}{\partial \mathcal{W}}\gamma - \left[-\hat{\boldsymbol{u}}\cdot\nabla\hat{\boldsymbol{u}} + \nu \nabla^2 \hat{\boldsymbol{u}} + \boldsymbol{f}\right]\right\Vert^2_2\mathrm{d}\boldsymbol{x}, 
\ee
The implementation and minimization of (\ref{WminNS2}) does not requires any special treatment and the projection, which is performed explicitly in fractional step methods, is automatically realized in EDNN by the least square solution of the linear system (\ref{OptCondApprox}) associated with (\ref{WminNS2}). The equivalence between (\ref{WminNS1}) and (\ref{WminNS2}) can be formally verified, 
\be
\begin{split}
    \nabla_{\gamma}\mathcal{J}   = & \left(\int_{\Omega}\frac{\partial \hat{\boldsymbol{u}}}{\partial \mathcal{W}}^{T}\frac{\partial \hat{\boldsymbol{u}}}{\partial \mathcal{W}}\mathrm{d}\boldsymbol{x}\right)\gamma_{opt} -\left(\int_{\Omega}\frac{\partial \hat{\boldsymbol{u}}}{\partial \mathcal{W}}^{T}\mathcal{N}_{\text{NS}}(\hat{\boldsymbol{u}})\mathrm{d}\boldsymbol{x}\right) \\
= &
    \left(\int_{\Omega}\frac{\partial \hat{\boldsymbol{u}}}{\partial \mathcal{W}}^{T}\frac{\partial \hat{\boldsymbol{u}}}{\partial \mathcal{W}}\mathrm{d}\boldsymbol{x}\right)\gamma_{opt} -\left(\int_{\Omega}\left(\mathcal{P}\frac{\partial \hat{\boldsymbol{u}}}{\partial \mathcal{W}}\right)^{T}\mathcal{N}_{\text{NS}}(\hat{\boldsymbol{u}})\mathrm{d}\boldsymbol{x}\right)\\
= &
    \left(\int_{\Omega}\frac{\partial \hat{\boldsymbol{u}}}{\partial \mathcal{W}}^{T}\frac{\partial \hat{\boldsymbol{u}}}{\partial \mathcal{W}}\mathrm{d}\boldsymbol{x}\right)\gamma_{opt} -\left(\int_{\Omega}\frac{\partial \hat{\boldsymbol{u}}}{\partial \mathcal{W}}^{T}\mathcal{P}^{T}\mathcal{N}_{\text{NS}}(\hat{\boldsymbol{u}})\mathrm{d}\boldsymbol{x}\right)\\
= &
    \left(\int_{\Omega}\frac{\partial \hat{\boldsymbol{u}}}{\partial \mathcal{W}}^{T}\frac{\partial \hat{\boldsymbol{u}}}{\partial \mathcal{W}}\mathrm{d}\boldsymbol{x}\right)\gamma_{opt} -\left(\int_{\Omega}\frac{\partial \hat{\boldsymbol{u}}}{\partial \mathcal{W}}^{T}\mathcal{P}\mathcal{N}_{\text{NS}}(\hat{\boldsymbol{u}})\mathrm{d}\boldsymbol{x}\right) = \nabla_{\gamma}\mathcal{J}_{P}\\
\end{split}
\ee
where $\mathcal{N}_{\text{NS}} = -\hat{\boldsymbol{u}}\cdot\nabla\hat{\boldsymbol{u}} + \nu \nabla^2 \hat{\boldsymbol{u}} + \boldsymbol{f}$ is the RHS of Navier-Stokes equation (\ref{NS2}) without the projection operator $\mathcal{P}$. The second equality above holds because the columns of $\partial \hat{\boldsymbol{u}}/\partial \mathcal{W}$ are all divergence-free, and the fourth equality uses the fact that $\mathcal{P}$ is an orthogonal projection operator. 
This validity an accuracy of this approach will also be demonstrated empirically through comparison of EDNN and analytical solutions of the incompressible Navier-Stokes equation.

\subsubsection{Taylor-Green vortex}
Two-dimensional Taylor-Green vortices are an exact time-dependent solution of the Navier-Stokes equations. This flow has been adopted extensively as a benchmark to demonstrate accuracy of various algorithms. The initial condition is, 
\be
\begin{split}
    u(x,y,t=0) &= U_0\mathrm{cos}\left(\frac{x}{L_x}\right)\mathrm{sin}\left(\frac{y}{L_y}\right)\\
    v(x,y,t=0) &= -U_0\mathrm{sin}\left(\frac{x}{L_x}\right)\mathrm{cos}\left(\frac{y}{L_y}\right), \\
\end{split}
\ee
and in absence of external forcing ($\mathbf{f}=0$) the time-dependent velocity field is, 
\be
\begin{split}
    u(x,y,t=0) &= U_0\mathrm{cos}\left(\frac{x}{L_x}\right)\mathrm{sin}\left(\frac{y}{L_y}\right)\mathrm{e}^{-2\nu t}\\
    v(x,y,t=0) &= -U_0\mathrm{sin}\left(\frac{x}{L_x}\right)\mathrm{cos}\left(\frac{y}{L_y}\right)\mathrm{e}^{-2\nu t}, \\
\end{split}
\ee
where $L_x = L_y = 2\pi$ are the dimensions of the flow domain. Periodicity is enforced on the boundaries of the domain. 

\begin{table}[b!]
    \caption{Parameters for the numerical solution of Taylor-Green Vortex using EDNN}
    \begin{center}\label{TGV_params}
    \begin{tabular}{ cccccc } 
     \hline
     Case & $L$ & $n_L$ & $N_x$ & $N_y$ & $\Delta t$ \\ 
     \hline
     1t & \multirow{9}{*}{$4$}  &  \multirow{4}{*}{$10$}& \multirow{4}{*}{$33$} & \multirow{4}{*}{$33$} & $1\times 10^{-2}$ \\ 
     2t &                       &                      &                       &  & $1\times 10^{-3}$ \\ 
     3t &                       &                       &                       &  & $1\times 10^{-4}$ \\ 
     4t &                       &                       &                       &  & $1\times 10^{-5}$ \\ \cline{3-6}
     5t &                       &  \multirow{4}{*}{$20$}& \multirow{4}{*}{$65$} & \multirow{4}{*}{$65$} & $1\times 10^{-2}$ \\
     6t &                       &                       &                       &  & $1\times 10^{-3}$ \\
     7t &                       &                       &                       &  & $1\times 10^{-4}$ \\
     8t &                       &                       &                       &  & $1\times 10^{-5}$ \\ \cline{3-6}
     9t &                       &       $30$         &            $129$      & $129$ &$1 \times 10^{-4}$ \\ 
    \hline
    \end{tabular}
    \label{TGV_params}
    \end{center}
\end{table}

\begin{figure}\label{TGV_sol2}
    \centering
    \begin{subfigure}{.25\linewidth}
    \centering
    \includegraphics[width=1.0\textwidth]{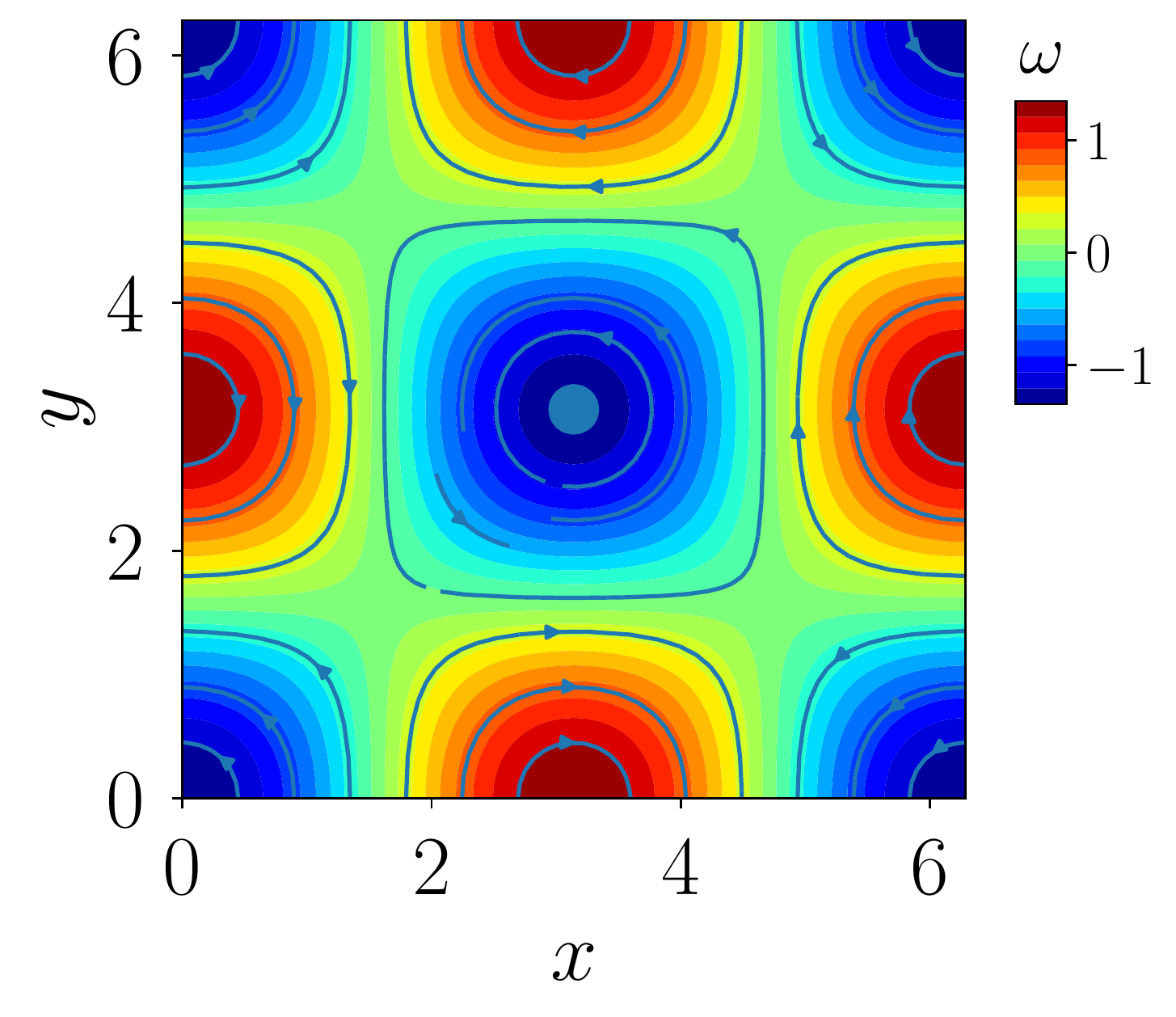}
    \caption{}
    \end{subfigure}%
    \begin{subfigure}{.25\linewidth}
    \centering
    \includegraphics[width=1.0\textwidth]{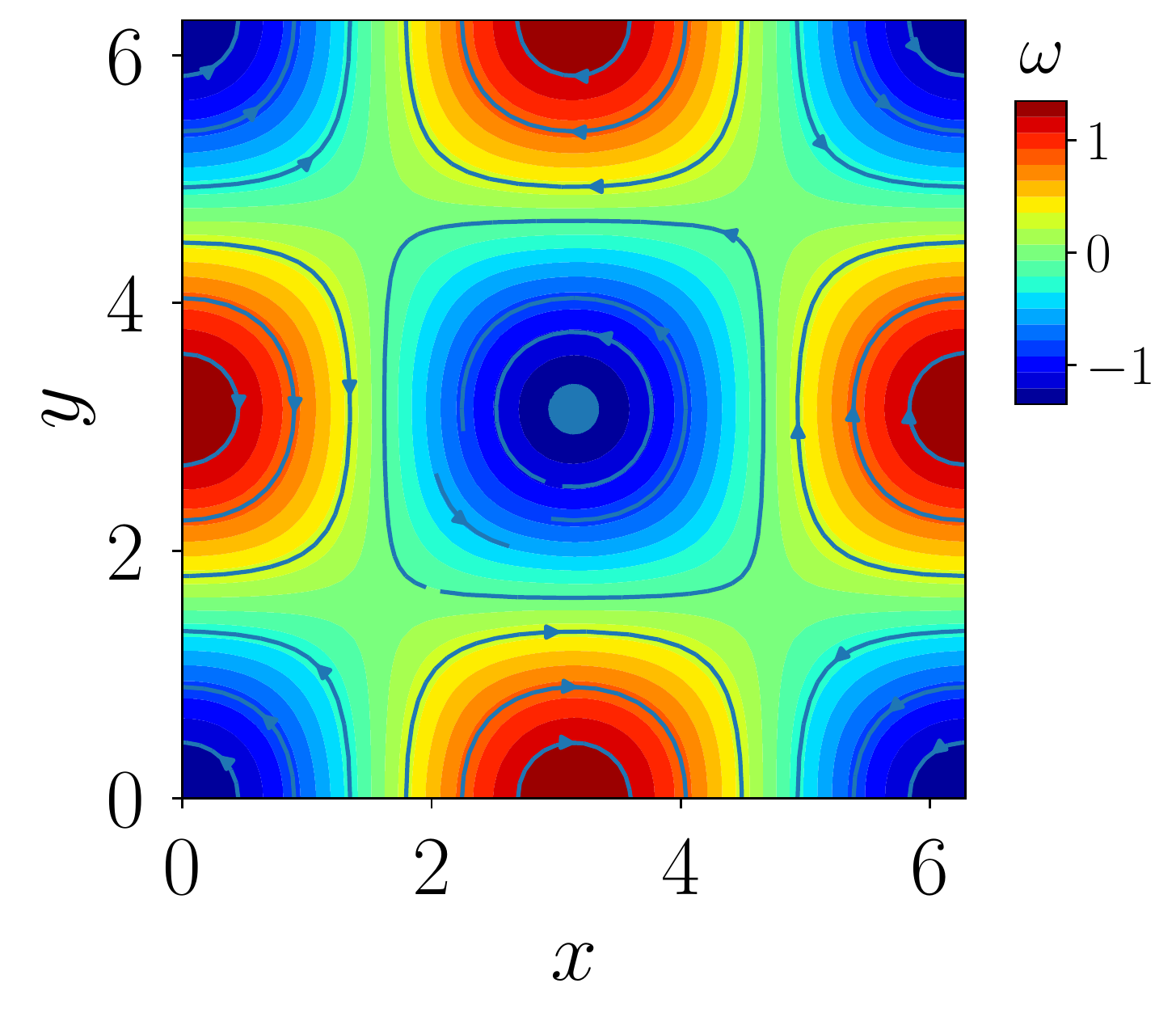}
    \caption{}
    \end{subfigure}
    \caption{Analytical and EDNN solution of Taylor-Green vortex at $t=0.2$. Color contours show the vorticity, and lines are the streamfunction. (a) Analytical solution. (b) Case 6t using EDNN. }
    \label{TGV_sol2}
\end{figure}

A comparison of the analytical and EDNN solutions is provided in figure \ref{TGV_sol2}.  The contours show the vorticity field $\omega = \nabla \times \mathbf{u} $ and lines mark streamlines that are tangent to the velocity field.  The prediction by EDNN shows excellent agreement with the analytical solution at $t=0.2$, and satisfies the periodic boundary condition.

In order to quantify the accuracy of EDNN predictions, a series of nine test cases, denoted 1t through 9t, were performed and are listed in Table \ref{TGV_params}.  All EDNN architectures are comprised of $L=4$ layers, and three network sizes were achieved by increasing the number of neurons per layer $n_L = \{10, 20, 30\}$.  The three values of $n_L$ were adopted for three resolutions of the solution points $(N_x, N_y)$ in the two-dimensional domain, and at each spatial resolution a number of time-steps $\Delta t$ were examined.

Quantitative assessment of the accuracy of EDNN is provided in figure (\ref{TGV_sol3}). First, the decay of the domain-averaged energy of the vortex $\mathcal{E} = (1/\vert\Omega\vert) \int_\Omega \mathbf{u}^2 d\Omega$ is plotted in panel (a) for all nine cases which all compare favorably to the analytical solution. The time-averaged root-mean-squared errors in the solution,
\be
    \epsilon = \frac{1}{T}\int_{0}^{T}\frac{\left\Vert u(t) - \hat{u}(t)\right\Vert_2}{\left\Vert u(t)\right\Vert_2}\mathrm{d}t
\ee
are plotted in panel (b).  
For any of the time-steps considered, as the number of solution points ($N_x,N_y$) is increased, and with it the number of neurons per layer $n_L$, the errors in the EDNN prediction is reduced.  In addition, as the time-step is reduced from $\Delta t = 10^{-2}$ to $10^{-4}$, the errors monotonically decrease.  Below $\Delta t = 10^{-4}$, the error saturates which is in part due to errors in the representation of the initial condition and from spatial discretization using the neural network. 
We have also verified that the solution satisfies the divergence-free condition to machine precision, which is anticipated because of the contraint was embedded in the EDNN design and derivatives are computed using automatic differentiation.


\begin{figure}\label{TGV_sol3}
\centering
\begin{subfigure}{.25\linewidth}
\centering
\includegraphics[width=1.0\textwidth]{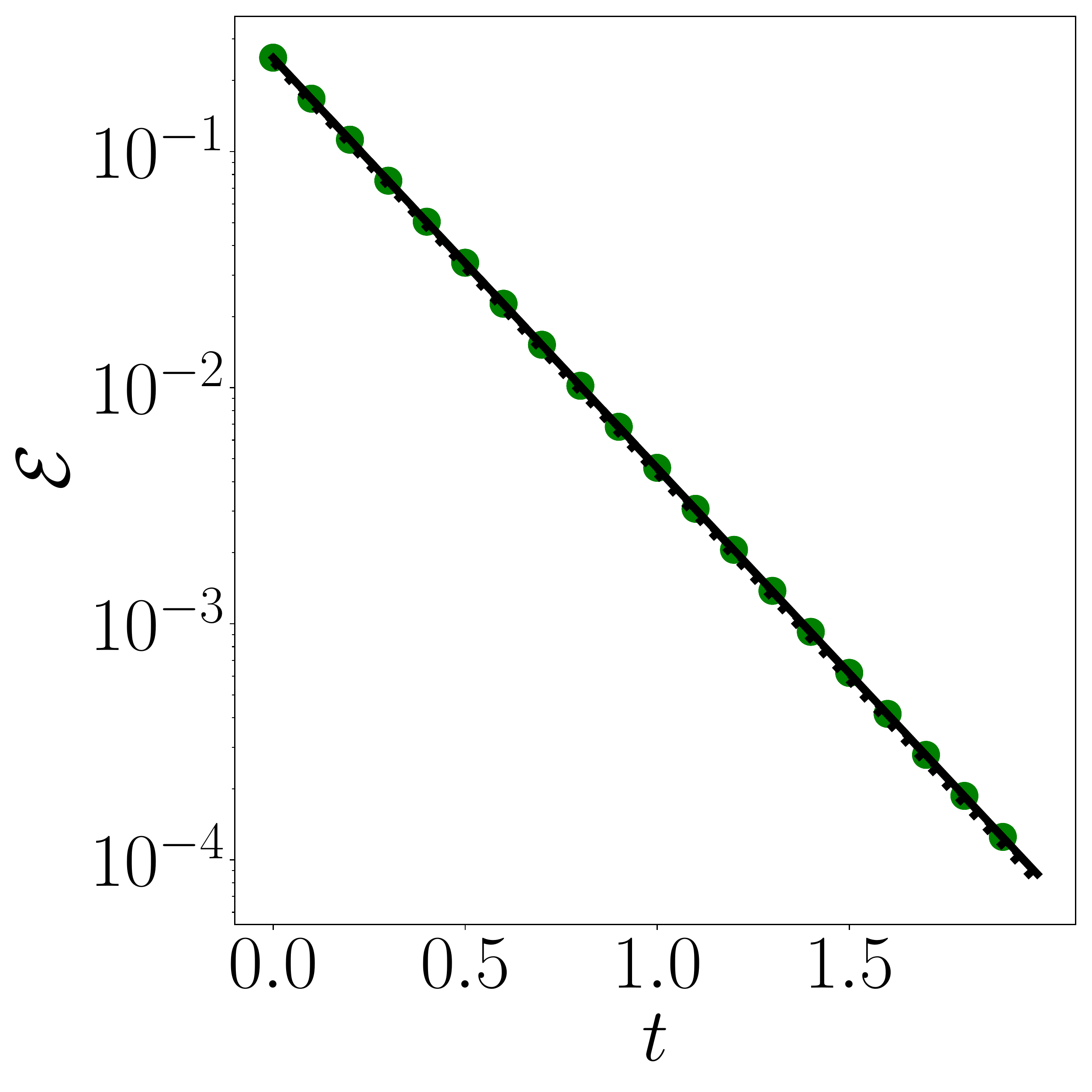}
\caption{}
\end{subfigure}%
\hspace*{24pt}
\begin{subfigure}{.25\linewidth}
\centering
\includegraphics[width=1.0\textwidth]{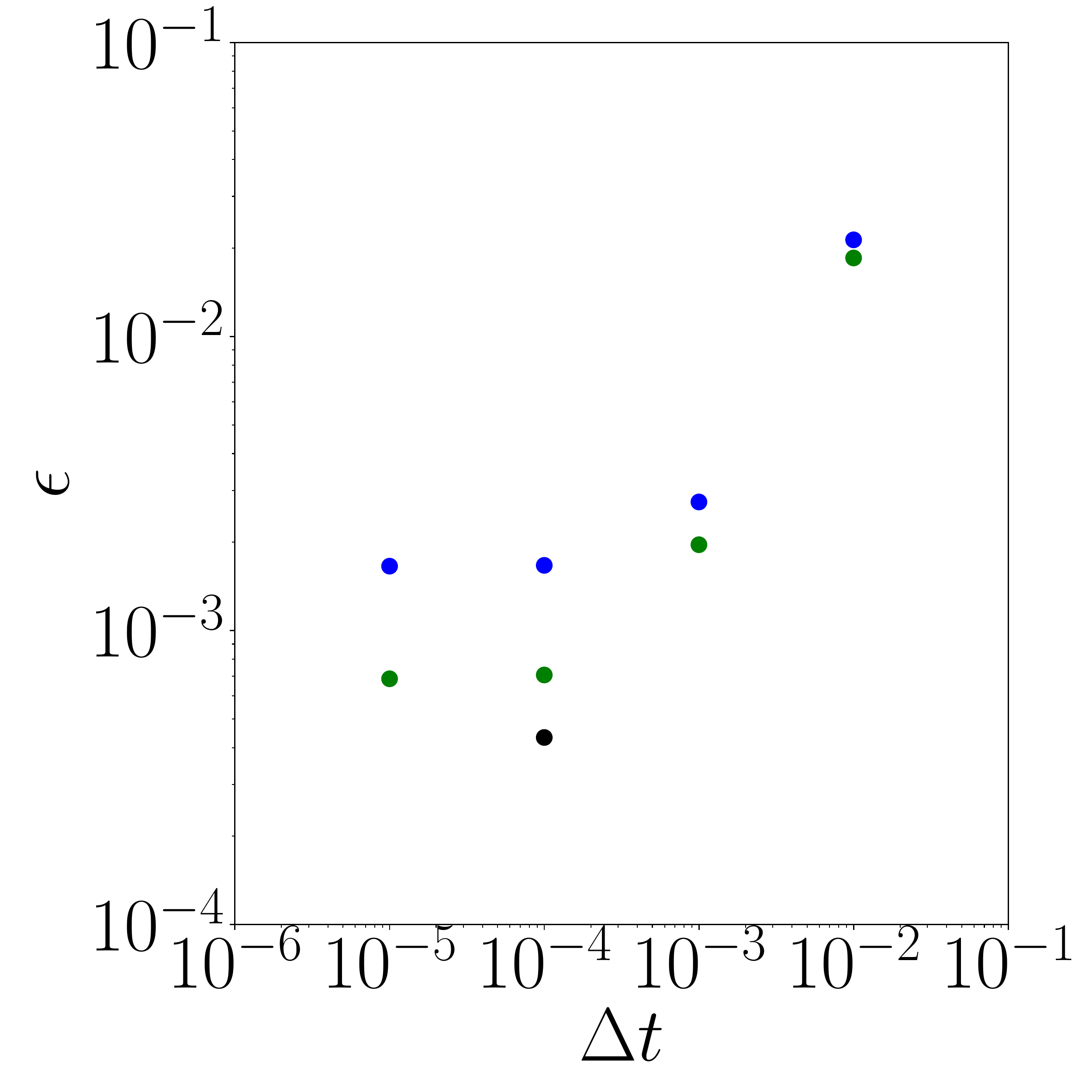}
\caption{}
\end{subfigure}
\caption{Quantitative assessment of EDNN solution for Taylor Green vortex. (a) Decay of Kinetic energy from EDNN and analytical solutions. (b) Relative error in EDNN prediction versus the time-step $\Delta t$. \mycircle{blue!30!black}: cases 1t to 4t;  \mycircle{green!30!black}: cases 5t to 8t; \mycircle{black!30!black}: case 9t.}
\label{TGV_sol3}
\end{figure}



\subsubsection{Kolmogorov flow}\label{Res: Kolmogorov flow}

The final Navier-Stokes example that we consider is the Kolmogorov flow, which is a low dimensional chaotic dynamical system that exhibits complex behaviors including instability, bifurcation, periodic orbits and turbulence\cite{chandler2013invariant, lucas2015recurrent}. The accurate simulation of long time chaotic dynamical system is important and also a challenge to the algorithm, thus we choose it as a numerical example.

Our objective here will be to demonstrate that EDNN can accurately predict trajectories of this flow in state space when starting from a laminar initial condition, and also long-time statistics when the initial condition is within the statistically stationary chaotic regime.  The latter objective is extremely challenging because very long-time integration is required for convergence of statistics, and 
will be demonstrated here using EDNN.  

The incompressible NS equation equations (\ref{NS1}) are solved with forcing in the horizontal $x$ direction, $\boldsymbol{f} = \chi \mathrm{sin}(ny)\mathbf{e}_x$ where $\chi = 0.1$ is the forcing amplitude and $n$ is the vertical wavenumber.  
Simulations starting from a laminar condition adopted the initial field, 
\be\label{Kflow_Init}
    \begin{split}
    u(x,y,t=0) &= 0\\
    v(x,y,t=0) &= -\mathrm{sin}\left(x\right), \\
    \end{split}
\ee
The spatial domain of the Kolmogorov flow is fixed on $[-\pi,\pi]^2$. The Reynolds number is defined as $\mathrm{Re}=\sqrt{\chi}/\nu$ consistent with  \cite{chandler2013invariant}. 
Independent simulations were performed using Fourier spectral discretization of the Navier-Stokes equations (see Table \ref{Kflow_params}), at high spectral resolution and with a small time-step because these are intended as reference solutions. Two forcing wavenumbers were considered: Case 1kfS with $n=4$ generates a laminar flow trajectory starting from equation (\ref{Kflow_Init}); Case 2kfs with $n=2$ adds random noise to the initial field (\ref{Kflow_Init}) in order to promote transition to a chaotic turbulent state, and flow statistics are evaluated once statistical stationarity is achieved.

\begin{table}\label{Kflow_params}
\caption{Parameters for Kolmogorov flow simulations using Fourier spectral methods and EDNN. }
\begin{center}
\begin{tabular}{ cccccccccc } 
 \hline
 & Case & $L$ & $n_L$  & $N_x$ & $N_y$ & $\Delta t$ & $\mathrm{Re}$ & $n$ & I.C.\\ 
 \hline
\multirow{2}{*}{Spectral} & 1kfS & & &  \multirow{2}{*}{$128$} & \multirow{2}{*}{$128$} & \multirow{2}{*}{$1\times 10^{-3}$} & \multirow{2}{*}{$33$} & 4 & L \\ \cline{9-10}
& 2kfS &  &                     &                       &  &  & &$2$ & T\\ 
\hline
\multirow{2}{*}{EDNN} & 1kfE & \multirow{2}{*}{$4$}  &  \multirow{2}{*}{$20$}& \multirow{2}{*}{$65$} & \multirow{2}{*}{$65$} & \multirow{2}{*}{$1\times 10^{-2}$} & \multirow{2}{*}{$33$} & 4 & L \\ \cline{9-10}
 & 2kfE &                       &                       &                       &                       &  &  &2 & T \\ 
\hline
\end{tabular}
\end{center}
\label{Kflow_params}
\end{table}

\begin{figure}\label{Kflow_laminar} 
    \centering
    \begin{subfigure}{.25\linewidth}
    \centering
    \includegraphics[width=1.0\textwidth]{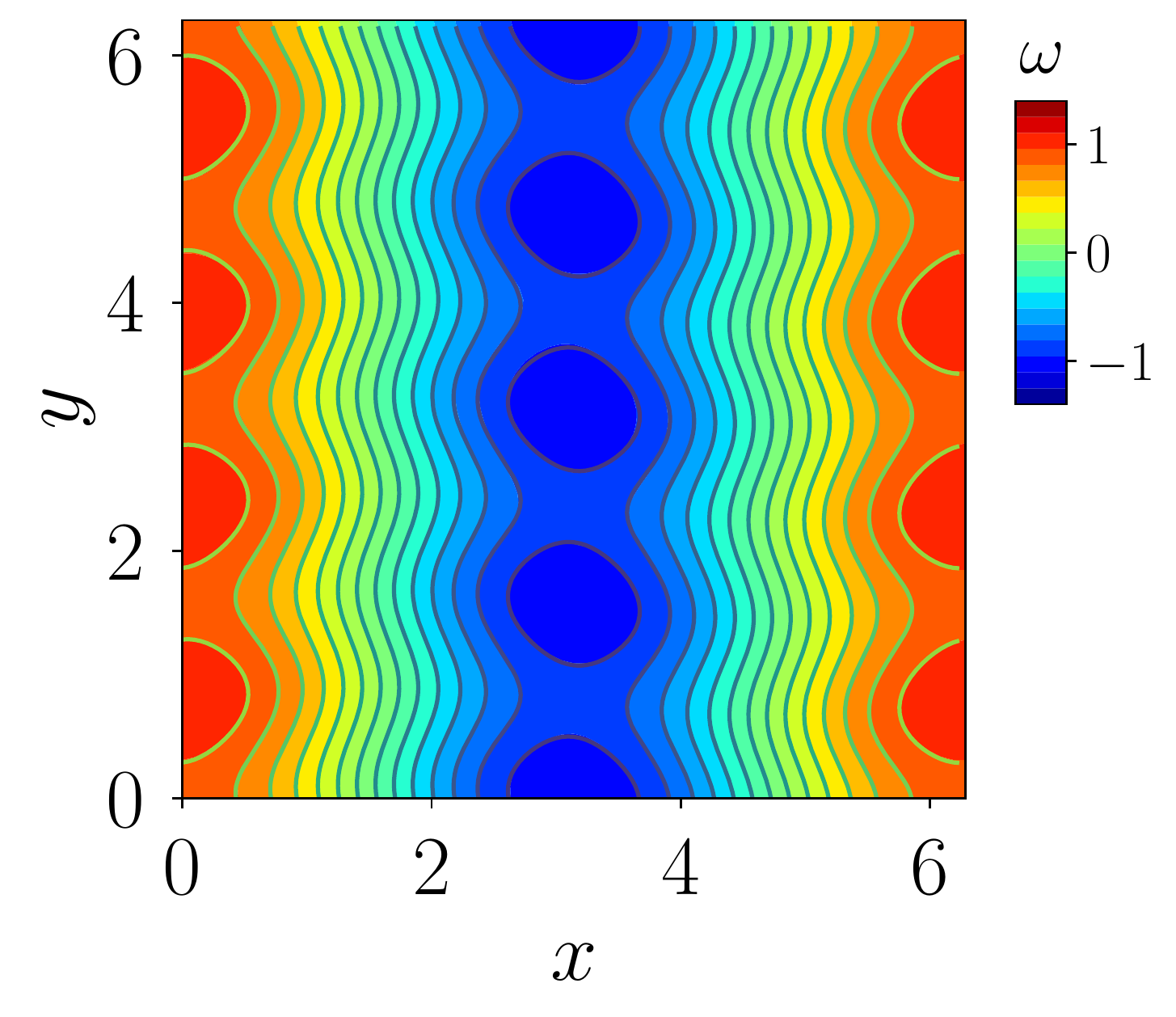}
    \caption{t=0.25}
    \end{subfigure}%
    \begin{subfigure}{.25\linewidth}
    \centering
    \includegraphics[width=1.0\textwidth]{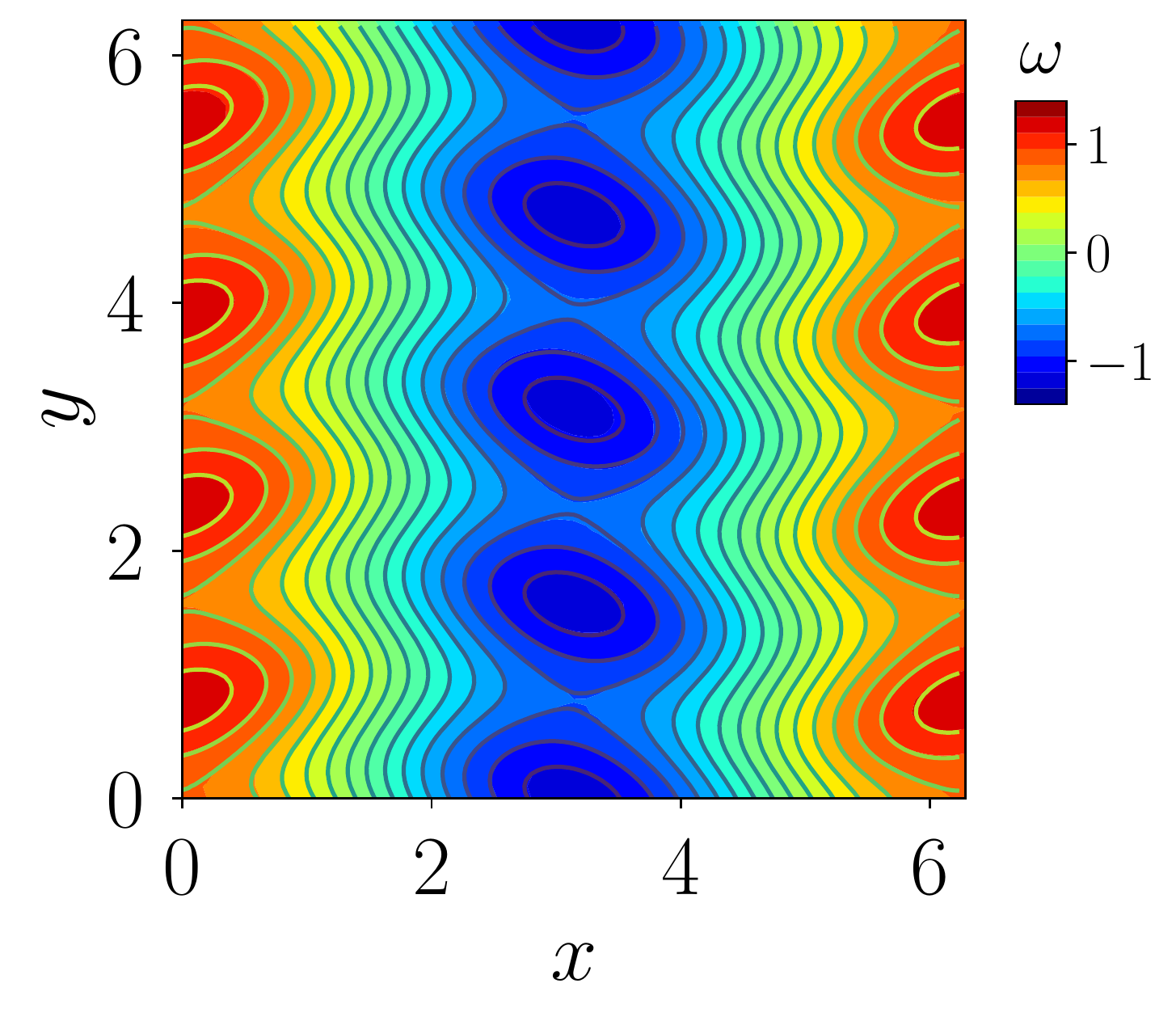}
    \caption{t=0.50}
    \end{subfigure}%
    \begin{subfigure}{.25\linewidth}
    \centering
    \includegraphics[width=1.0\textwidth]{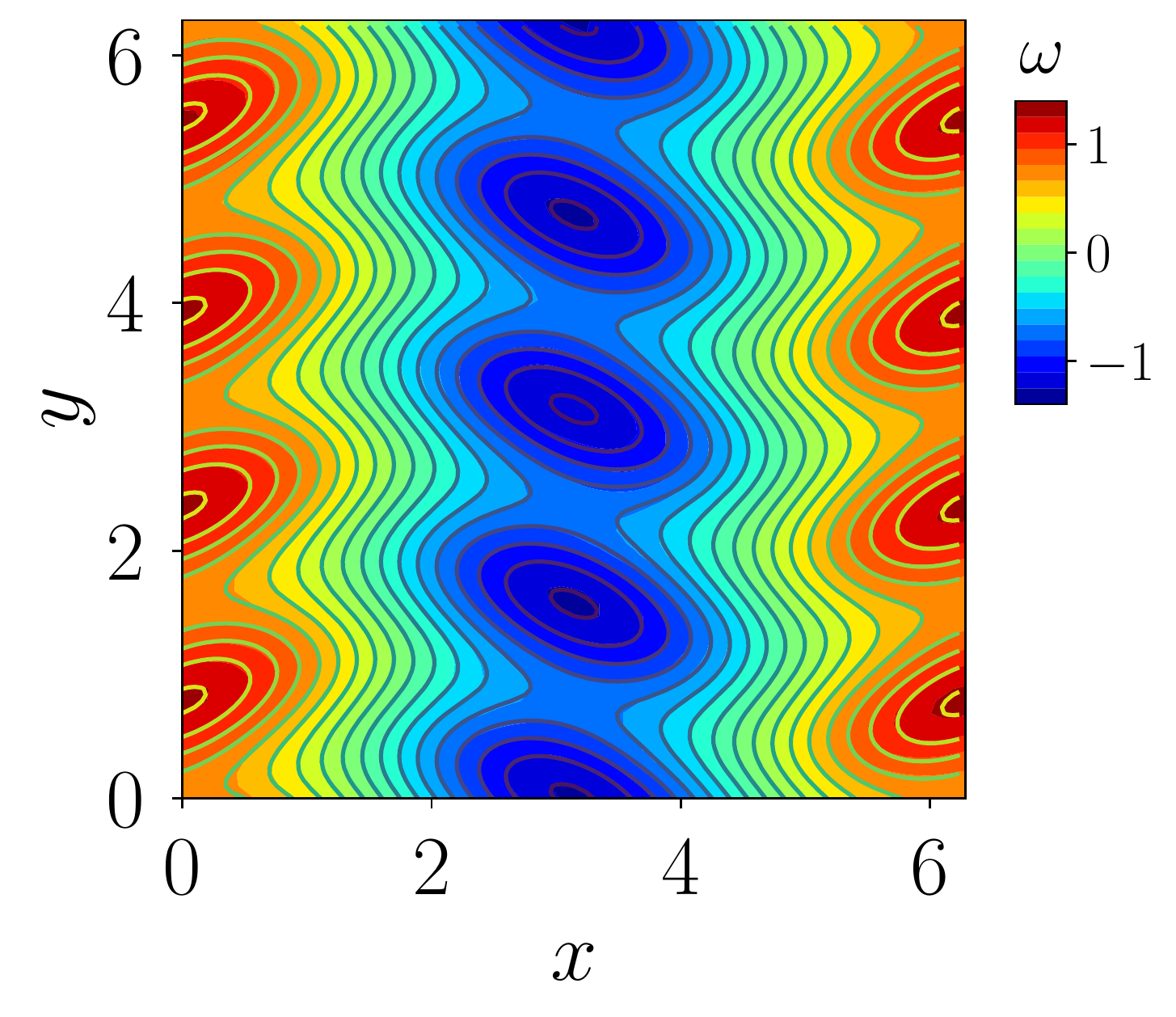}
    \caption{t=0.75}
    \end{subfigure}%
    \begin{subfigure}{.25\linewidth}
    \centering
    \includegraphics[width=1.0\textwidth]{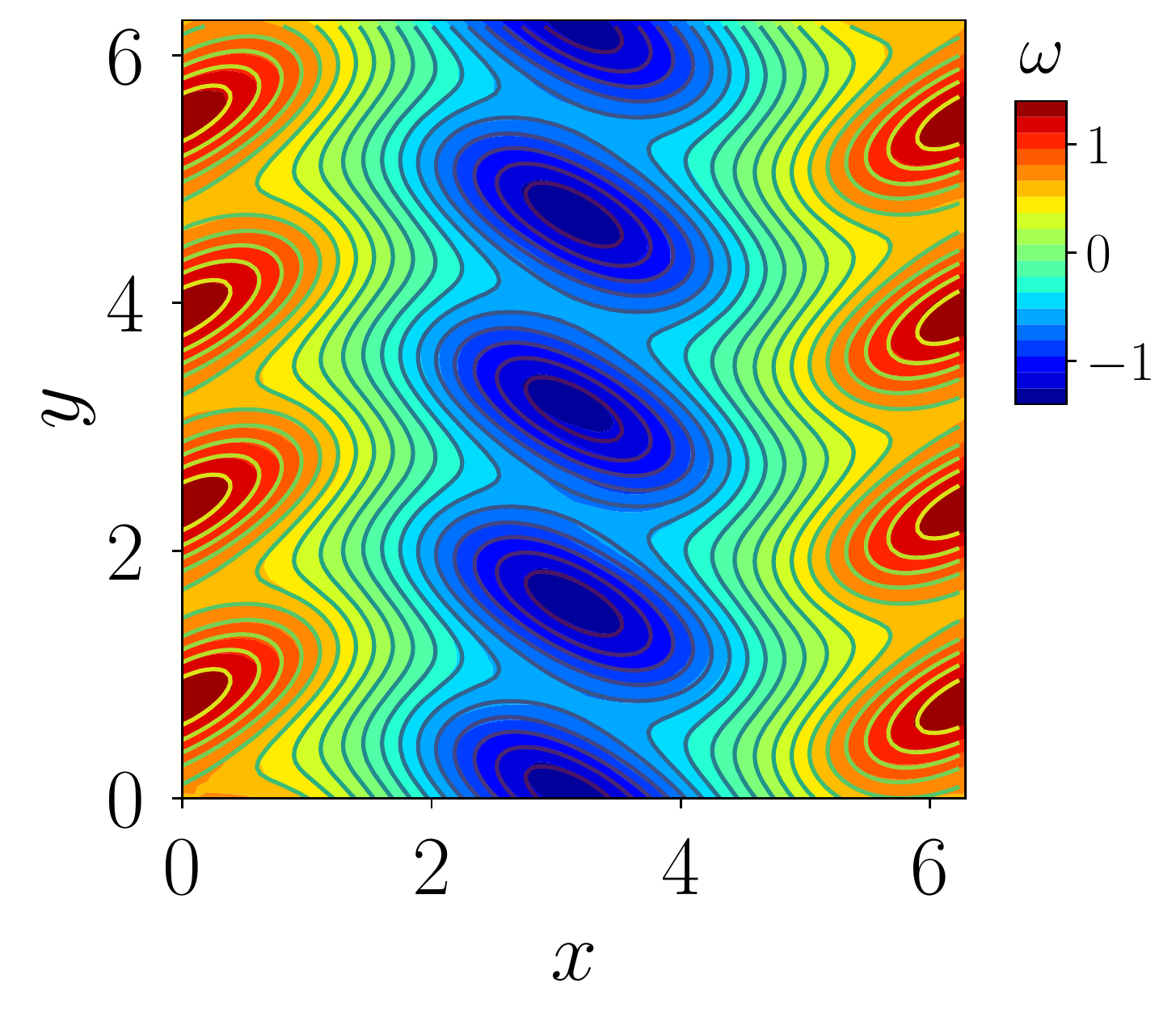}
    \caption{t=1.0}
    \end{subfigure}%
    \caption{Comparison of instantaneous vorticity $\omega$ in Kolmogorov flow using EDNN and spectral method. Colors contours are from case 1kfE (EDNN) and line contours are from 1kfS (spectral). }
\label{Kflow_laminar}
\end{figure}


\begin{figure}\label{Kflow_chaotic} 
    \centering
    %
    \begin{subfigure}{.24\linewidth}
    \centering
    \includegraphics[width=1.0\textwidth]{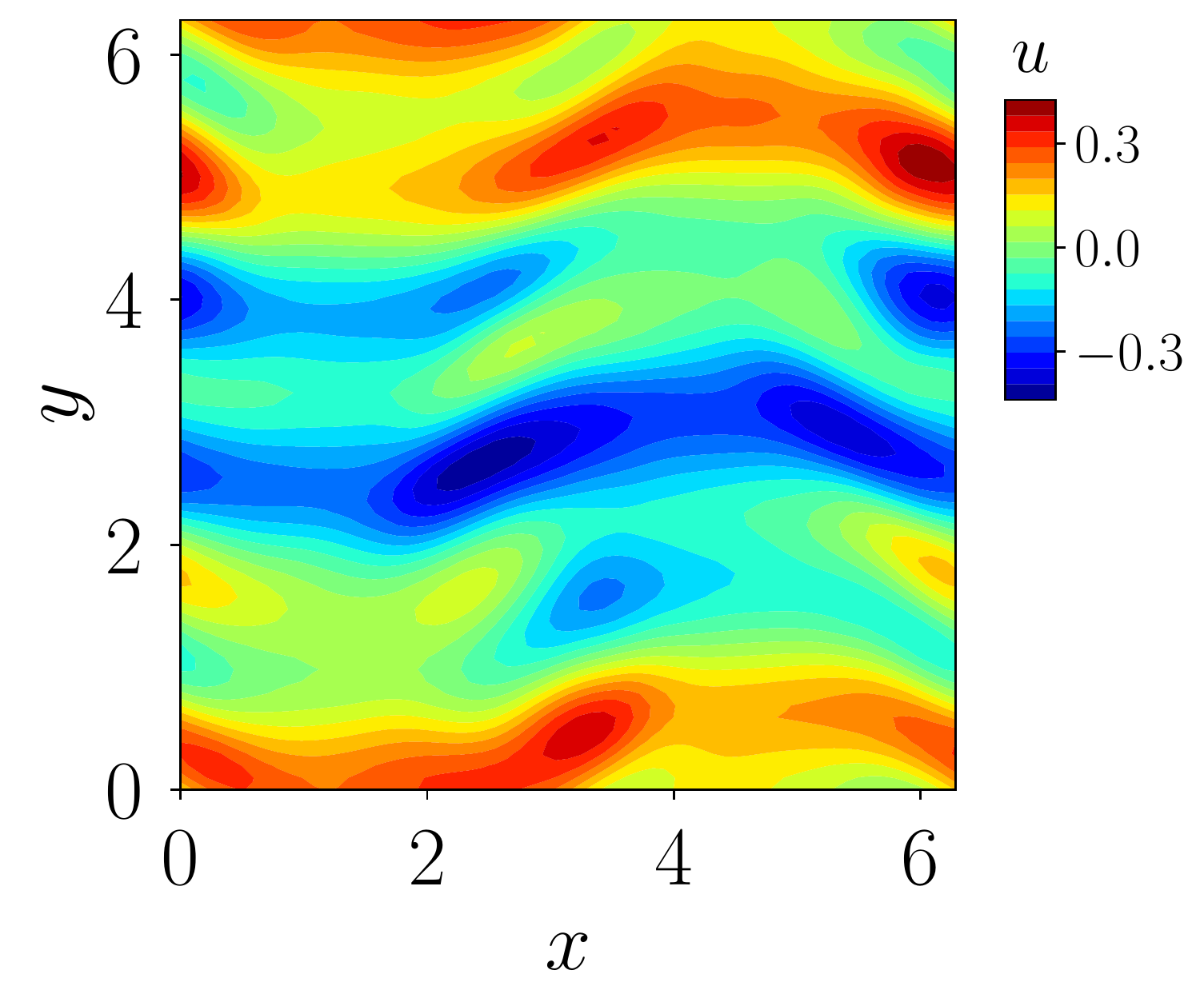}
    \caption{n=4, $u$}
    \end{subfigure}%
    \begin{subfigure}{.24\linewidth}
    \centering
    \includegraphics[width=1.0\textwidth]{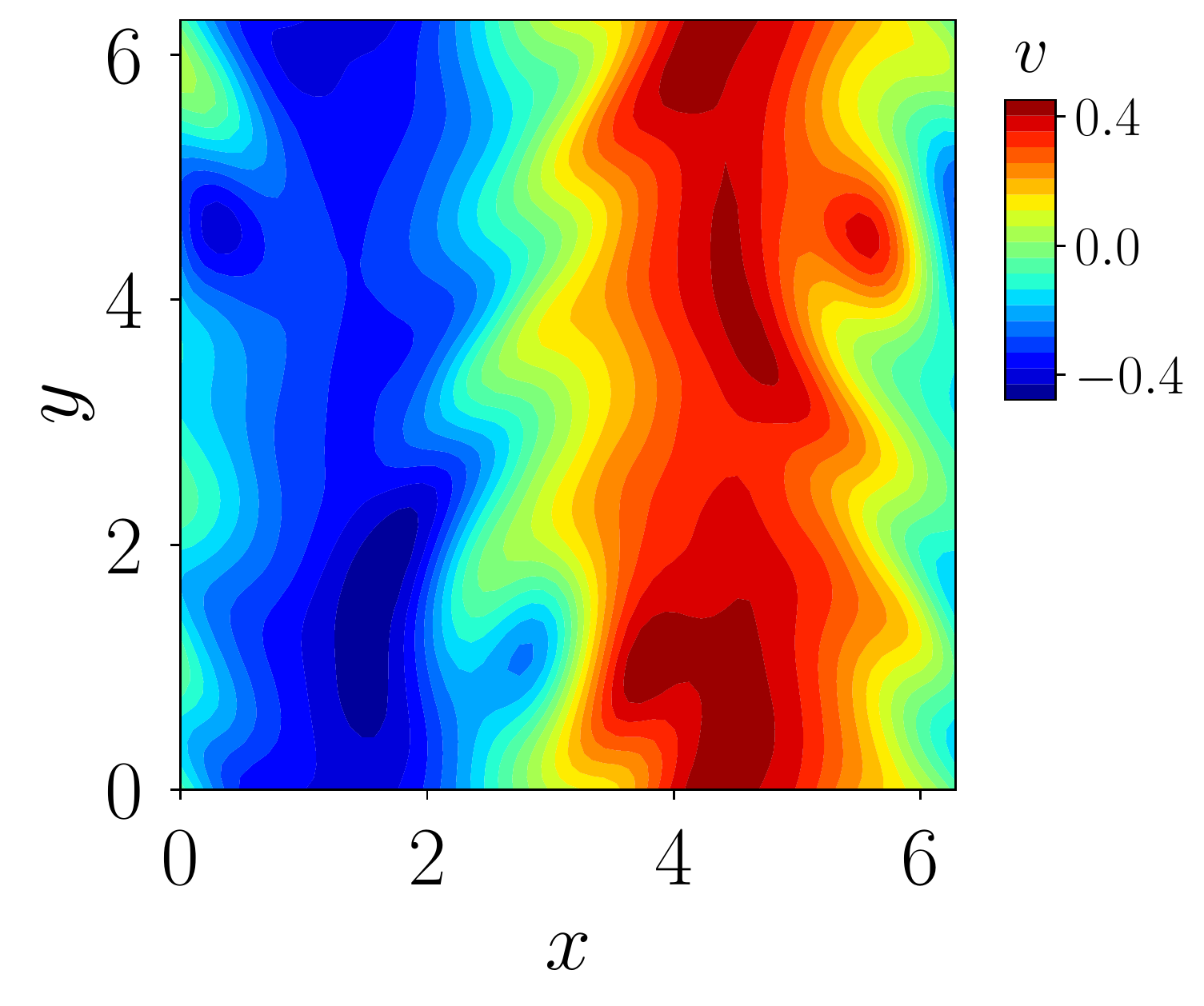}
    \caption{n=4, $v$}
    \end{subfigure}
    \caption{Instantaneous (a) horizontal and (b) vertical velocities in the turbulent state at $t = 10^{5}$ withe forcing wavenumber $n=4$, simulated using EDNN.}
    \label{Kflow_chaotic}
\end{figure}

The EDNN simulations parameters are also listed in Table \ref{Kflow_params}, all using the same network architecture, number of spatial points and time-step.  The laminar case (1kfE, $n=4$) shares the same initial condition (\ref{Kflow_Init}) as the spectral solution; The turbulent case (2kfE, $n=2$), on the other hand, was simulated starting from a statistically stationary state extracted from the spectral computation, and therefore statistics were evaluated immediately from the initial time.  

The laminar cases 1kfs and 1kfE are compared in figure \ref{Kflow_laminar}.  Contours of the vorticity field $\omega = \nabla \times \mathbf{u} $ are plotted using color for the EDNN solution and lines for the spectral reference case, and their agreement demonstrates the accuracy of EDNN in predicting the time evolution.
If noise is added to the initial condition, these cases transition to turbulence.  A snapshot of such turbulent velocity field obtained using EDNN at very long time, $t=10^{4}$, is shown in the figure \ref{Kflow_chaotic} to confirm that transition to turbulence can indeed be achieved.  It is well known, however, that convergence of first and second order statistics when $n=4$ is extremely challenging, and requires sampling over a duration on the order of at least $10^6$ time units \citep{lucas2015recurrent}. 
We therefore adopt $n=2$ for the computation of turbulent flow statistics, where convergence is achieved faster, but nonetheless still requiring long challenging integration times.  A realization of the statistically stationary state from EDNN (case 2kfE) is shown in figure \ref{Kflow_stats}. The velocity field shows evidence of the forcing wavenumber, but is clearly irregular.  
Long-time flow statistics from both EDNN and the spectral simulation (2kfs) also shown in the figure.  The black curves are the mean velocity and blue ones show the root-mean-squared perturbations as a function of the vertical coordinate.  Agreement of EDNN prediction with the reference spectral solution is notable, even though the spatio-temporal resolution in EDNN is coarser.  We also note that these simulations were performed over a very long times ($6\times 10^{5}$ for spectral and $4\times 10^{5}$ for EDNN).  Performing such long-time evolutions of turbulent trajectories has never been demonstrated with  
existing neural-network approaches,
and was here demonstrated to be accurately achieved with EDNN.



\begin{figure}\label{Kflow_stats}
    \centering
    \begin{subfigure}{.24\linewidth}
    \centering
    \includegraphics[width=1.0\textwidth]{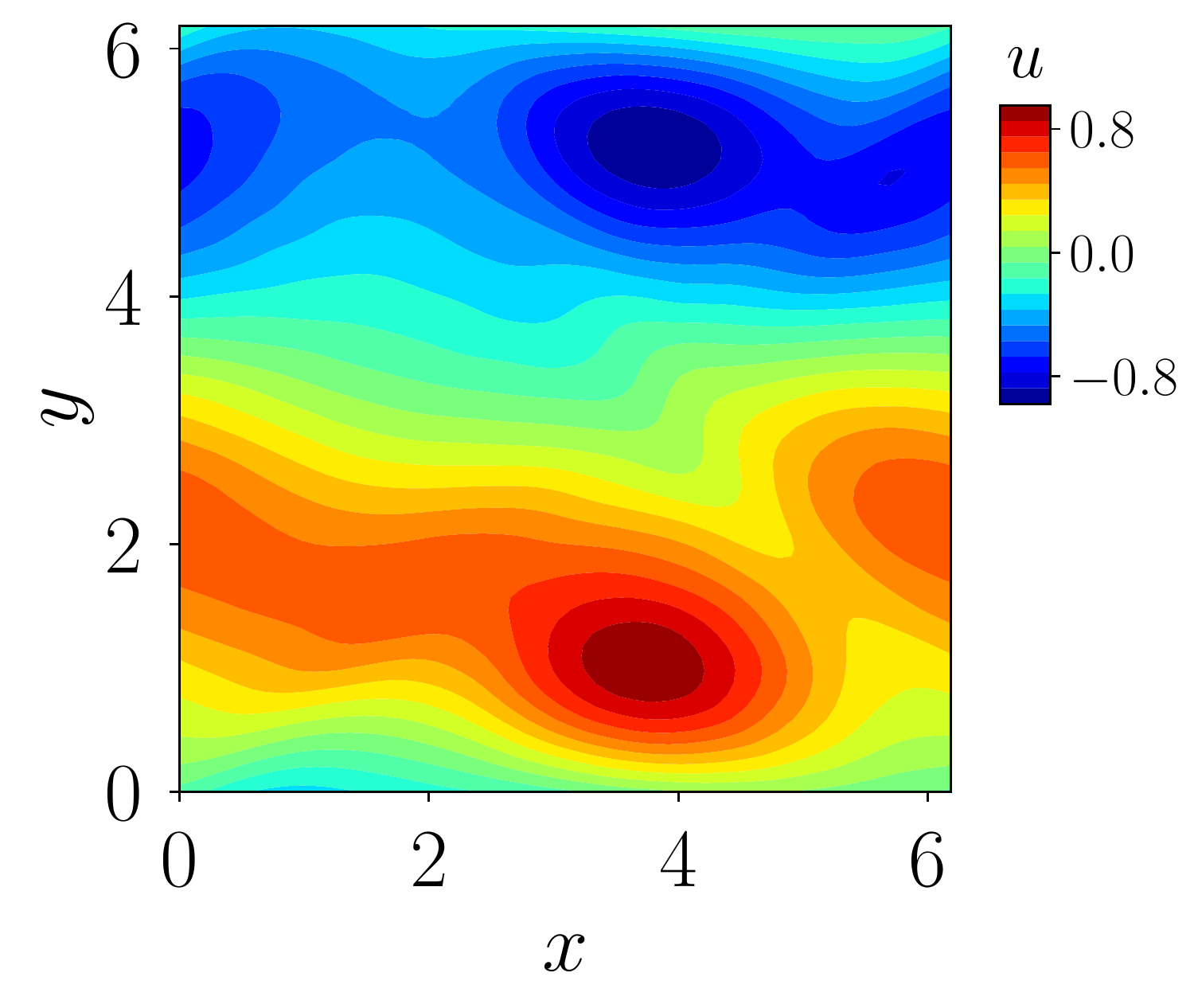}
    \caption{n=2, $u$}
    \end{subfigure}%
    \begin{subfigure}{.24\linewidth}
    \centering
    \includegraphics[width=1.0\textwidth]{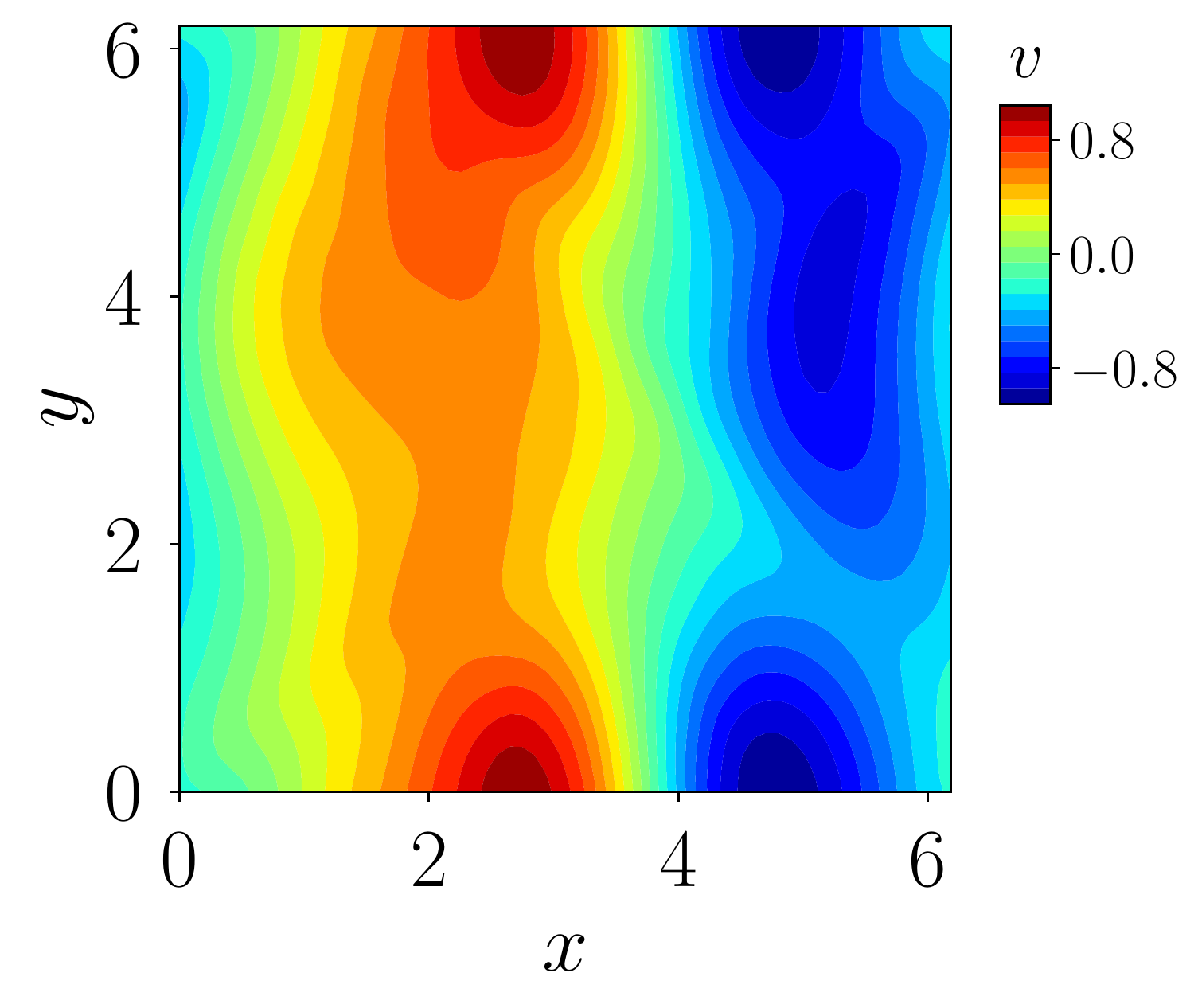}
    \caption{n=2, $v$}
    \end{subfigure}%
    \begin{subfigure}{.24\linewidth}
    \centering
    \includegraphics[width=1.0\textwidth]{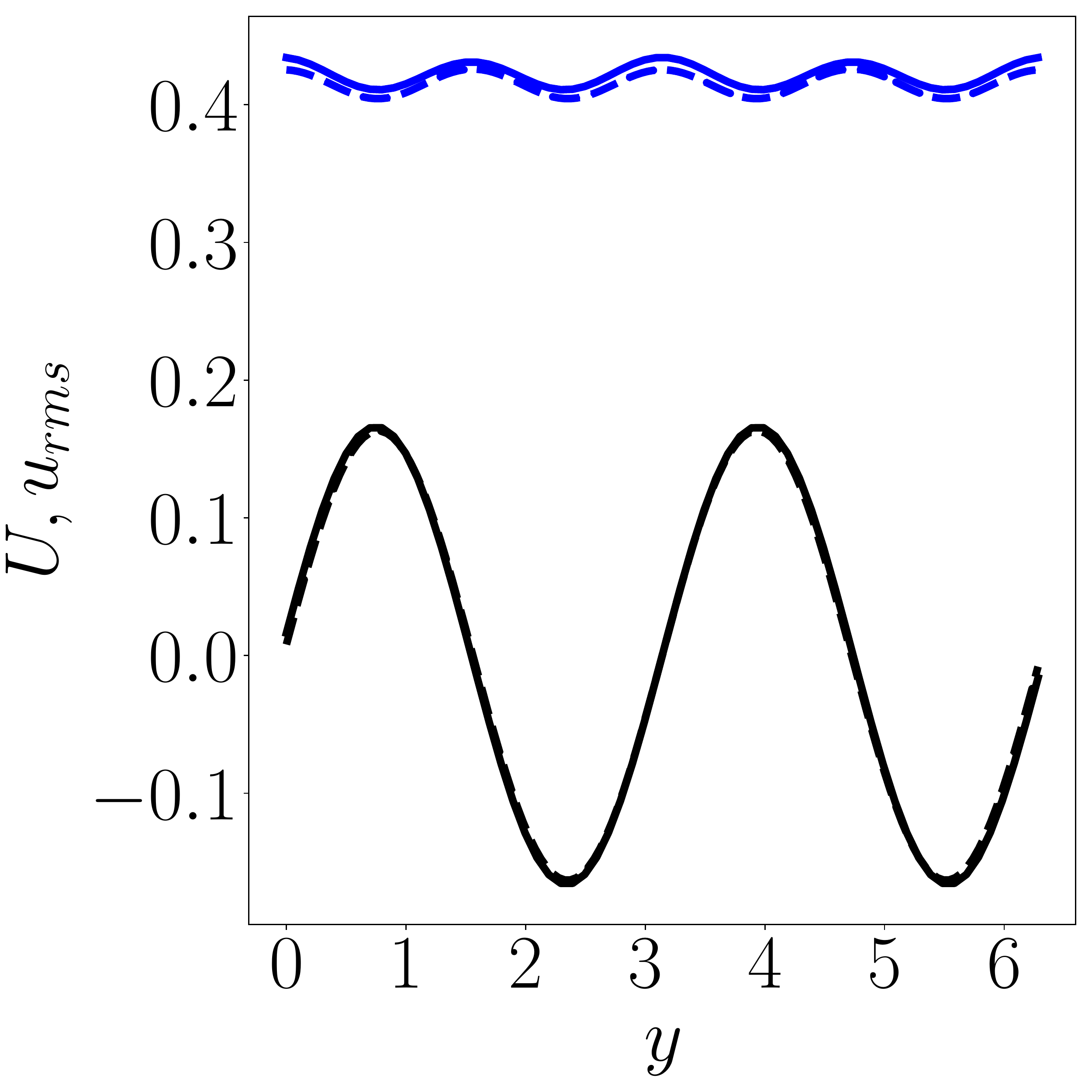}
    \caption{$U$, $u_{rms}$}
    \end{subfigure}%
    \begin{subfigure}{.24\linewidth}
    \centering
    \includegraphics[width=1.0\textwidth]{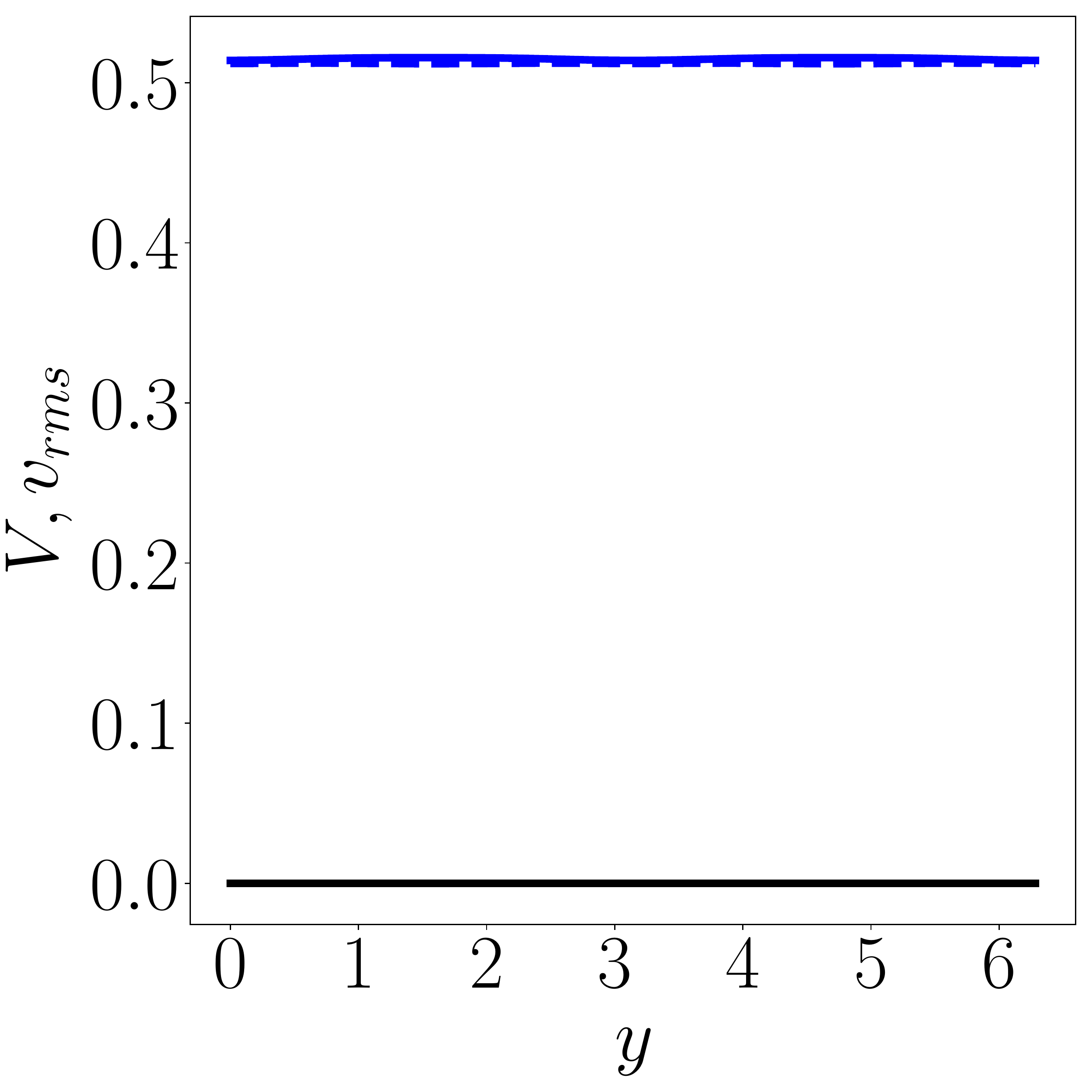}
    \caption{$V$, $v_{rms}$ }
    \end{subfigure}
    \caption{Instantaneous snapshots and long-time statistics of the chaotic Kolmogorov flow with forcing wavenumber $n=3$.  
    (a) horizontal and (b) vertical velocities in the turbulent state at $t=10^{5}$ simulated using EDNN.
    (c,d) Statistics of horizontal and vertical velocitiies, respectively, evaluated from spectral simulation (solid line, case 2kfs) and EDNN (dashed lines, case 2kfE).  
    Black lines are the mean velocities, and blue lines are the root-mean-squared fluctuations. }
    \label{Kflow_stats}
\end{figure}

\section{Conclusions}\label{Conclusions}
A new framework is introduced for simulating the evolution of solutions to partial differential equations using neural network. Spatial dimensions are discretized using the neural network, and automatic differentiation is used to compute spatial derivatives.  The temporal evolution is expressed in terms of an evolution equation for the network parameters, or weights, which are updated using a marching scheme.  Starting from the initial network state that represents the initial condition, the weights of the Evolutional Deep Neural Network (EDNN) are marched to predict the solution trajectory of the PDE over any time horizon of interest. Boundary conditions and other linear constraints on the solution of the PDE are enforced on the neural network by the introduction of auxiliary functions and auxiliary operators. 
The EDNN methodlogy is flexible, and can be easily adapted to other types of PDE problems.  For example, in boundary-layer flows, the governing equations are often marched in the parabolic streamwise direction \citep{cheung2010linear, cheung2011nonlinear, park2019sensitivity} .  In this case, the inputs to EDNN would be the spatial coordinates in the cross-flow plane, and the network weights would be marched in the streamwise direction instead of time.  

Several PDE problems were solved using EDNN in order to demonstrate its versatility and accuracy, including two-dimensional heat equation, linear wave equation and Burgers equation.  Tests with the Kuramoto-Sivashinsky equation focused on the ability of EDNN to accurately predict bifurcations.  For the two-dimensional incompressible Navier-Stokes equations, we introduced an approach where projection step which ensures solenoidal velocity fields is automatically realized by an embedded divergence-free constraints.  We then simulated decaying Taylor-Green vortices.   
In all cases, the solutions from EDNN show good agreement with either analytical solutions or reference spectral discretizations.  In addition, the accuracy of EDNN monotonically improves with the refinement of neural network structure, and the adopted spatio-temporal resolution for representing the solution.  
For Navier-Stokes equations, we also considered the evolution of Kolmogorov flow in the early laminar regime as well as its long-time statistics in the chaotic turbulent regime.  Again the predictions of EDNN were accurate, and its ability to simulate long time horizons was highlighted.

EDNN has several noteworthy characteristics.  Previous neural network methods for time dependent PDE
perform an optimization on the whole spatio-temporal domain. In contrast, the state of EDNN only represents an instantaneous snapshot of the PDE solution. Thus, the structural complexity of EDNN can be significantly smaller than 
other approaches for a specific PDE problem. Secondly, EDNN maintains explicit time dependency and causality, while most of other methods only try to minimize the penalty on equation residuals. Thirdly, EDNN can simulate very long time evolution of chaotic solutions of the PDE, which is difficult to achieve in other NN based methods. 

The main computational cost of EDNN involves automatic differentiation of the network outputs to evaluate the equation operator $\mathcal{N}_{\boldsymbol{x}}(\boldsymbol{u})$, the formation of the Jacobian matrix $\mathbf{J}$, and inverting the linear system $\mathbf{J}^T\mathbf{J}$. The key difference to conventional, structured finite-difference methods for example is that the linear system is not sparse which incurs computational cost. This relative weakness is outweighed by the flexibility of EDNN, where the method is simple to implement for any differential operator, complex geometric grids are not required and dynamic refinement of collocation points can be trivially performed during the evolution of the solution. 
The cost of solving the dense linear system can be mitigated in future work by domain decomposition: deploying small networks on sub-domains with interface boundary conditions (e.g.\,enforced using the approach in \S\ref{Method: Embedded constraints}) would lead to a block-sparse system matrix, and lends itself to parallelism for computational acceleration.  
Noteworthy is that for the incompressible Navier-Stokes equations, the EDNN design guarantees that the flow is divergence free without an explicit projection step that requires solution of a separate elliptic pressure equation.



\section*{Acknowledgements}
The authors are grateful to Prof.\,Charles Meneveau for his comments on an initial draft of this work.

\bibliography{citation}
\end{document}